\documentclass{aa}  

\usepackage{graphicx}
\usepackage{txfonts}
\usepackage{subfig}
\usepackage{pdflscape}
\usepackage{adjustbox}
\begin{document}

   \titlerunning{FUV-to-FIR view of LBGs at $z\sim3$}
   \authorrunning{\'Alvarez-M\'arquez, Burgarella, Buat et al.}

   \title{Rest-frame far-ultraviolet to far-infrared view of Lyman break galaxies at $z\sim3$: Templates and Dust attenuation}


\author{J.~\'Alvarez-M\'arquez\inst{1,2}
\and D.~Burgarella\inst{2}
\and V.~Buat\inst{2}
\and O.~Ilbert\inst{2}
\and P.~G.~P\'erez-Gonz\'alez\inst{1}
}

\institute{
$^{1}$
Centro de Astrobiolog\'ia (CSIC-INTA), Carretera de Ajalvir, 28850 Torrej\'on de Ardoz, Madrid, Spain\\
\email{javier.alvarez@cab.inta-csic.es}\\
$^{2}$
Aix-Marseille Universit\'e, CNRS, LAM (Laboratoire d'Astrophysique de Marseille) UMR7326, 13388, France\\Descargar
}

   \date{Received ; accepted }

 
  \abstract
{}
{This work explores, from a statistical point of view, the rest-frame far-ultraviolet (FUV) to far-infrared (FIR) emission of a population of Lyman-break galaxies (LBGs) at $z\sim3$ that cannot be individually detected from current FIR observations.}
{We performed a stacking analysis over a sample of $\sim$17000 LBGs at redshift $2.5<z<3.5$ in the COSMOS field. The sample is binned as a function of UV luminosity ($L_{\mathrm{FUV}}$), UV continuum slope ($\beta_{\mathrm{UV}}$), and stellar mass (M$_{*}$), and then stacked at optical ($BVriz$ bands), near-infrared ($YJHKs$ bands), IRAC (3.6, 4.5, 5.6 and 8.0 $\mu$m), MIPS (24$\mu$m), PACS (100 and 160~$\mu$m), SPIRE (250, 350, and 500~$\mu$m), and AzTEC (1.1mm) observations.  We obtained 30 rest-frame FUV-to-FIR spectral energy distributions (SEDs) of LBGs at $z\sim3$, and analyzed these with the CIGALE SED-fitting analysis code. We were able to derive fully consistent physical parameters, that is, M$_{*}$, $\beta_{\mathrm{UV}}$, $L_{\mathrm{FUV}}$, $L_{\mathrm{IR}}$, A$_{FUV}$, star formation rate, and the slope of the dust attenuation law; we built a semiempirical library of 30 rest-frame FUV-to-FIR stacked LBG SEDs as functions of $L_{\mathrm{FUV}}$, $\beta_{\mathrm{UV}}$, and M$_{*}$.}
{We used the so-called IR-excess ($IRX \equiv L_{\mathrm{IR}} / L_{\mathrm{FUV}}$) to investigate the dust attenuation as a function of $\beta_{\mathrm{UV}}$ and M$_{*}$. Our LBGs, averaged as a function of $\beta_{\mathrm{UV}}$, follow the well-known IRX-$\beta_{\mathrm{UV}}$ calibration of local starburst galaxies. Stacks as a function of $M_{*}$ follow the IRX-$M_{*}$ relationship presented in the literature at high $M_{*}$ ($\log$(M$_{*}$ [M$_{\odot}$])> 10). However, a large dispersion is shown in the IRX-$\beta_{\mathrm{UV}}$ and IRX-$M_{*}$ planes, in which the $\beta_{\mathrm{UV}}$ and M$_{*}$ are combined to average the sample. Additionally, the SED-fitting analysis results provide a diversity of dust attenuation curve along the LBG sample, and their slopes are well correlated with M$_{*}$. Steeper dust attenuation curves than Calzetti's are favored in low stellar mass LBGs ($\log(M_{*} [M_{\odot}]) < 10.25$), while grayer dust attenuation curves are favored in high stellar mass LBGs ($\log(M_{*} [M_{\odot}]) > 10.25$). We also demonstrate that the slope of the dust attenuation curves is one of the main drivers that shapes the IRX-$\beta_{\mathrm{UV}}$ plane.}
{}


   \maketitle

\section{Introduction}\label{introduction}

About 20 years ago, the \textit{Hubble Space Telescope} (HST) allied to the 10 m class ground-based telescopes opened up a window on the first 2 Gyr of cosmic times \citep{Madau1996}. Lyman-break galaxies (LBGs) represent the largest sample of star-forming galaxies known at high redshift ($z>2.5$) owing to the efficiency of their selection technique in deep broadband imaging survey (Lyman-break or dropout technique, \citealt{steidel96}). These galaxies have been a key population to investigate the mass assembly of galaxies during the first gigayears of the universe \citep{Shapley01, Somerville01, Madau1996, steidel96, Giavalisco2002, Blaizot2004, Shapley2005, Baugh05, Verma2007, Magdis2008, Stark2009, ChapmanCasey2009, LoFaro2009, Magdis2010a, Pentericci2010, Oteo2013a, Bian2013, Bouwens2015, Oesch2015, Roberts-Borsani2016, Stefanon2017, Oesch2018}.

The rest-frame ultraviolet (UV) light, mostly emitted by young and massive stars, has been commonly used as a star formation rate (SFR) tracer. However, the interstellar dust scatters or absorbs the UV light, and hence only a fraction of the energy output from star formation is observed in the UV. The rest is re-emitted in the infrared (IR) by the heated dust. The general picture of star formation activity across cosmic time peaks around $z\sim2-4$ and significantly drops at $z>4$ (see \citealt{Madau2014} for a review). The obscured star formation dominates the total SFR density (SFRD) over the redshift range $0<z<3$, and corresponds to half of the SFRD at $z=3.6$ \citep{Burgarella2013}. Then, it is necessary to combine the UV and IR emission to determine the complete energy budget of star formation. However, because of the limited sensitivity of far-infrared (FIR) and radio observations, most of the current information at redshifts $z>3-4$ are obtained from UV observations of LBG samples that need to be corrected for dust attenuation (e.g. \citealt{Madau2014}). 

Only a few LBGs have been directly detected at $z\sim3$ in the mid-infrared (MIR) and FIR \citep{Magdis2012,Casey2012,Oteo2013} thanks to \textit{Spitzer} and \textit{Herschel}. This sample is biased against submillimeter bright galaxies and is not representative of the LBG population in terms of stellar mass (M$_{*}$), dust attenuation, and SFR \citep{Burgarella2011,Oteo2013}. The recently gain in sensitivity with the Atacama Large Millimeter Array (ALMA) at submillimeter and millimeter wavelengths provides some insight into the obscured SFR at high-$z$ \citep{Capak2015,Bouwens2016,Fudamoto2017,McLure2018}. Even so, it is still very difficult to obtain large samples to carry out statistical and representative analyses. Therefore, stacking analysis techniques \citep{Dole2006} have been applied to relatively large samples of LBGs to derive their FIR/submillimeter emission \citep{Magdis2010b, Magdis2010c, Rigopoulou2010, Coppin2015, Alvarez-marquez2016, Koprowski2018}.

Owing to the lack of FIR/submillimeter information of individual LBGs, empirical recipes are used to correct the UV emission for dust attenuation. The most commonly adopted is the relation between the UV continuum slope ($\beta_{\mathrm{UV}}$, \citealt{Calzetti1994}) and the so-called IR excess ($IRX \equiv L_{\mathrm{IR}} / L_{\mathrm{FUV}}$) calibrated on local starburst galaxies by \citeauthor{Meurer1999} (M99 hereafter; 1999). Despite its general use to infer the dust attenuation at high-$z$, there are several complications that make this method uncertain. The $\beta_{\mathrm{UV}}$ is known to be sensitive to the intrinsic UV spectra of galaxies, which depends on the metallicity (Z), age of the stellar population, and star formation history (SFH), and the shape of the dust attenuation curve. Stacking analyses of LBG and star-forming galaxies at redshifts of $1.5 < z < 5$ have shown different behaviors; some of these results follow the M99 relation \citep{Magdis2010c,Reddy2012,Koprowski2018, McLure2018} \textbf{and} others lie above \citep{Coppin2015, Bourne2017} or below this relation \citep{Alvarez-marquez2016, Bouwens2016, Reddy2018}. These deviations have been shown to be driven by the stellar masses \citep{Alvarez-marquez2016, Bourne2017}, the shape of the dust attenuation curve \citep{Salmon2016, LoFaro2017}, and the sample selection \citep{Buat2015}. Direct detections of individual LBGs and IR-selected sources have suggested a large scatter in the IRX-$\beta_{UV}$ plane (e.g., \citealt{Oteo2013,Casey2015,Fudamoto2017_det}). In addition, the stellar mass has been found to correlate with the dust attenuation, and its relation does not seem to evolve with redshift \citep{Reddy2010,Buat2012,Heinis2014,Whitaker2014,Pannella2015,Alvarez-marquez2016,Bouwens2016,Whitaker2017} at least in the intermediate range $1.0 < z < 4.0$.

With the objective to investigate deeply the dust attenuation in LBGs at high-z, a sample of 22,000 LBGs in the redshift range $2.5<z<3.5$ were selected in the COSMOS field \citep{Scoville2007} and presented in a first publication (\citealt{Alvarez-marquez2016}, AM16 hereafter). A statistically controlled stacking analysis from the FIR to millimeter wavelengths (100 $\mu$m to 1.1mm) was applied to derive their full IR spectral energy distribution (SEDs). Thanks to the large LBG sample, the stacking analysis was performed in different subsamples as a function of several parameters: $L_{\mathrm{FUV}}$, $\beta_{\mathrm{UV}}$, and M$_{*}$. We investigated the dust properties for each subsample of LBGs and studied their evolution in the IRX-$\beta_{UV}$, IRX-M$_{*}$, and IRX-$L_{\mathrm{FUV}}$ planes. 

In this paper, we extend the initial work by applying a new stacking analysis from optical to millimeter wavelengths. This approach allows us to obtain the full rest-frame far-ultraviolet (FUV) to FIR SEDs of LBGs at $z\sim3$, and derive the main physical parameters by mean of SED-fitting analysis techniques. This allows us to study the amount of dust attenuation and the shape of the dust attenuation curve that best represents the LBG population at $z\sim3$. The outline of this paper is as follows. In Section \ref{data}, we describe the COSMOS field data and photometry used in our analysis. In Section \ref{sample}, the definition of the LBG sample and its associated photometric redshift, M$_{*}$, $L_{\mathrm{FUV}}$, and $\beta_{UV}$ are presented. In Section \ref{stacking_analysis}, we define the method to stack the LBG sample from optical to FIR observations and obtain the stacked rest-frame FUV-to-FIR LBGs SEDs at $z\sim3$. In Section \ref{sed_fitting}, a SED-fitting analysis with CIGALE SED-fitting analysis code is performed on the stacked LBGs SED at $z\sim3$ to derive their physical properties and build a library of semiempirical templates of LBGs at $z\sim3$. In Section \ref{att}, we discuss the dust attenuation results obtained for the LBG sample in terms of the IRX and which dust attenuation curve is more likely to be representative for LBGs at $z\sim3$. Finally, Section \ref{conclusions} presents the summary and conclusions. 

Throughout this paper we use a standard cosmology with $\Omega_{\rm m} = 0.3$, $\Omega_{\Lambda} = 0.7$, and Hubble constant $H_{0} = 70$ km s$^{-1}$ Mpc$^{-1}$, and the AB magnitude system. When comparing our data to others studies, we assume no conversion is needed for stellar mass estimates between \cite{Kroupa2001} and \cite{Chabrier2003} IMFs. To convert from \cite{Salpeter1955} to \cite{Chabrier2003} IMFs, we divide $M_{*}$ $_{\mathrm Salpeter}$ by 1.74 \citep{Ilbert2010}.


\section{Data}\label{data}

We used the available optical to FIR imaging from the COSMOS field, and the optical/NIR multicolor catalog (\citealt{Capak2007}, version 2.0). The specific data sets are presented in the following Section.

\subsection{Optical and near-infrared data}\label{data_optical}

We used optical broadband imaging (B$_{\mathrm J}$, V$_{\mathrm J}$, r$^{+}$, i$^{+}$, z$^{++}$) from the COSMOS-20 survey \citep{Capak2007, Taniguchi2007, Taniguchi2015}, observed at the prime-focus camera (Suprime-Cam) on the 8.2 m Subaru Telescope. In combination with the YJHK$_{s}$ broadband imaging released in the UltraVISTA DR2\footnote{\url{www.eso.org/sci/observing/phase3/data_release/uvista_dr2.pdf}} \citep{McCracken2012} and observed with the \textit{VISTA InfraRed CAMera} (VIRCAM) instrument on the \textit{Visible and Infrared Survey Telescope for Astronomy} (VISTA). The original images have been homogenized to the same point spread function (PSF) with a full width at half maximum (FWHM) equal to 0.8$^{\prime\prime}$ by \citeauthor{Laigle2016} (2016, private communication). 

We additionally used optical/NIR broadband photometry (Subaru: $B_{\mathrm J}$, $V_{\mathrm J}$, $g^{+}$, $r^{+}$, $i^{+}$, $z^{+}$, and VISTA: $Y$,$ J$, $H$ and $K_{s}$) from the COSMOS multicolor catalog (\citealt{Capak2007}, version 2.0). The photometry was performed using SExtractor in dual-image mode over an aperture of 3$^{\prime\prime}$ centered at the position of the $i^{+}$ band detection. The COSMOS multicolor catalog was updated by \cite{Laigle2016}, in which they provide a new NIR selection catalog on the ultra VISTA-DR2 observations. However, we still used the i-band selection catalog (\citealt{Capak2007}, version 2.0) because the LBG sample, used in this analysis (see Sect. \ref{sample}), has been proposed to be detected at V$_{\mathrm J}$ and i$^{+}$ bands (AM16).

\subsection{Mid- and far-infrared imaging}\label{midfarinfrared}

We use the \textit{Infrared Array Camera} (IRAC: 3.6, 4.5, 5.8, and 8.0 $\mu$m) and the \textit{Multiband Imaging Photometer for SIRTF} (MIPS: 24 $\mu$m) observations from the SCOSMOS survey \citep{Sanders2007} as part of the \textit{Spitzer} Cycle 2 and 3 Legacy Programs. The IRAC and MIPS observations present a 5$\sigma$ flux sensitivity of 0.9, 1.7, 11.3, 14.9, and 80 $\mu$Jy, respectively. 

\textit{Herschel} Space Observatory \citep{Pilbratt2010} mapped the COSMOS field at 100 and 160 with the \textit{Photodetector Array Camera and Spectrometer} (PACS;  \citealt{Poglitsch2010}), and 250, 350, and 500~$\mu$m with the \textit{Photometric Imaging Receiver} (SPIRE; \citealt{Griffin2010}) as part of the \textit{Evolutionary Probe survey} (PEP; \citealt{Lutz2011}) and \textit{Herschel Multi-Tiered Extragalactic Survey} (HerMES, \citealt{Oliver2012}). The PACS images (100 and 160~$\mu$m) present point-source sensitivities of 1.5 mJy and 3.3 mJy and a PSF FWHM of 6.8$^{\prime\prime}$ and 11$^{\prime\prime}$. The SPIRE maps (250, 350, and 500~$\mu$m) have PSF FWHM of 18.2$^{\prime\prime}$, 24.9$^{\prime\prime}$, and 36.3$^{\prime\prime}$, 1$\sigma$ instrumental noise of 1.6, 1.3, and 1.9 mJy~beam$^{-1}$, and 1$\sigma$ confusion noise of 5.8, 6.3, and 6.8 mJy~beam$^{-1}$ \citep{Nguyen2010}. These maps were downloaded from HeDaM\footnote{Herschel Database in Marseille: \url{http://hedam.lam.fr/HerMES/}}


\section{LBG sample}\label{sample}

The LBG sample for this analysis is a subsample of that selected and characterized by AM16. The original LBG sample was selected in the COSMOS field by means of the classical U-dropout technique \citep{steidel96} using the broadband filters $u^{*}$, $V_{\mathrm J}$, and $i^{+}$. This original sample contains $\sim$22000 LBGs in the redshift range $2.5 < z < 3.5$ and UV luminosities $\log(L_{\mathrm{FUV}}[L_{\odot}]) > 10.2$. 

One of the objectives in this analysis is to perform a consistent rest-frame FUV to FIR stacking analysis. However, the available UltraVISTA survey ($YJHK_{s}$ bands) covers a reduced area of the COSMOS field of 1.5 deg$^{2}$. To be fully consistent and stack the same LBGs at any wavelength, we restricted the original LBGs sample to the limited area given by the UltraVISTA survey. Therefore, our final LBGs sample contains $\sim$17000 LBGs, which are all LBG from the original sample enclosed in the UltraVISTA area.  

\subsection{Photometric redshift and stellar mass}\label{photometricdata}

We used the photometric redshifts (photo-$z$) and stellar masses computed by \citeauthor{Ilbert2009} (2009, version 2.0) for i-band detected sources in the COSMOS field. The photo-$z$ in the range $1.5 < z < 4$ were tested against the zCOSMOS faint sample and faint DEIMOS spectra, showing an accuracy of $\sim$3\% (\citeauthor{Ilbert2009} 2009, version 2.0). The stellar masses were derived by applying SED-fitting techniques to the available optical and NIR photometry, and assuming; \cite{BruzualCharlot2003} single stellar population (SSP), exponentially declining SFH, and \cite{Chabrier2003} initial mass function (IMF). \cite{Ilbert2010} suggested that the i-band photometric redshift catalog is 90\% complete at 5$\mu$Jy and 50\% complete at 1$\mu$Jy in the IRAC 3.6 $\mu$m images.

\subsection{Definition of $L_{\mathrm{FUV}}$ and $\beta_{\mathrm{UV}}$ for each LBG}\label{slope_uvluminosity}

The UV continuum slope, $\beta_{\mathrm{UV}}$, is defined following the method presented by \cite{Finkelstein2012}. They performed a rest-frame UV to optical SED-fitting analysis and obtained the $\beta_{\mathrm{UV}}$ by fitting a power law to the derived best-fit synthetic spectral model. In our case, we limited our LBG SEDs to the available optical-to-NIR broadband photometry (optical: $B_{\mathrm J}$, $V_{\mathrm J}$, $g^{+}$, $r^{+}$, $i^{+}$, $z^{+}$ and NIR: $Y$,$ J$, $H$ and $K_{s}$) that lie at the rest-frame wavelength range $1000 < \lambda_{rest-frame}~[\AA] < 3500$. This allows us to perform an SED-fitting analysis over the same rest-frame spectral range for each LBGs independent of its redshift. 

We used CIGALE (Code Investigating GALaxy Emission; \citealt{Burgarella2005, Noll2009, Boquien2019}, see Sect. \ref{sed_fitting}) to perform the rest-frame UV-to-optical SED-fitting analysis and derive the best-fit model spectrum for each LBG of the sample. We assume \cite{BruzualCharlot2003} synthetic stellar population libraries and \cite{Chabrier2003} IMF. We varied the metallicity (0.02 < Z [Z$_{\odot}$]< 1.00 ), age of the stellar population (1 Myr < t < t$_{H}$), dust extinction (0 < A$_{V}$ < 2 mag, using the \cite{Calzetti2000} dust attenuation law), and SFH (SFH $\propto$ e$^{-t/\tau}$, where $\tau = $ 0.1, 10, 100, 10$^{3}$, 10$^{4}$, 10$^{5}$, -300, -10$^{3}$, -10$^{4}$ Myr).

The value $\beta_{\mathrm{UV}}$ is calculated by power-law fit ($f_{\lambda} \propto \lambda^{\beta_{\mathrm{UV}}}$) to the best-fit synthetic stellar population model that has been derived by CIGALE. The power-law fit is performed at rest-frame wavelength range $1250 < \lambda_{rest-frame}~[\AA]  < 2000$ following the spectral windows defined by \cite{Calzetti1994}. We excluded the range 2000  < $\lambda$ [$\AA$]< 2600 for two reasons: i) to omit the dust feature at 2175 $\AA$ and ii) to have a homogeneous rest-frame wavelength range independent of the redshift of the galaxy. The value $L_{\mathrm{FUV}}$ is computed in a squared bandpass defined as the GALEX FUV band ($\lambda_{eff}$ = 152.8 nm and effective bandwidth = 11.4 nm). We checked the uncertainties in the $\beta_{\mathrm{UV}}$ and $L_{\mathrm{FUV}}$ estimations using the mock analysis tool from CIGALE (see Sect. \ref{physical_param_sed_fitting} for details of the mock analysis tool). We obtained a dispersion in the $\beta_{\mathrm{UV}}$ of $\Delta\beta_{\mathrm{UV}}\sim0.2$ and a very robust $L_{\mathrm{FUV}}$ measure with $\Delta\log(L_{\mathrm{FUV}}[L_{\odot}])\sim0.03$. Additionally, we obtained a systematic offset between the $\beta_{\mathrm{UV}}$ calculated by power-law fit directly to the photometry (AM16) and that derived from power-law fit to the best-fit model obtained by SED-fitting analysis of the photometry (see Appendix \ref{beta_comparison} for a detailed discussion).  


\section{Stacking analysis}\label{stacking_analysis}

Stacking analysis is a technique to determine the mean flux density of a population of sources that are individually too dim to be detected in a confusion-limited maps (e.g., \citealt{Dole2006,Marsden2009,Bethermin2012,Heinis2013}). Our LBG sample is complete at $V_{\mathrm J}$ and $i^{+}$ bands, where the sample has been selected by means of the U-dropout technique. However, the detection of individual LBGs at other optical/NIR bands are dropped up to 70\% of completeness, and it gets worse at longer wavelengths as far as only few of the LBGs could be detected at FIR observations (SPIRE 250 $\mu$m). We then decided to average the signals of multiple LBGs from optical-to-FIR observations to obtain fully consistent mean rest-frame FUV-to-FIR SEDs of LBGs at $z\sim3$. This allows us to detect statistically the LBG population at redder wavelength at the expense of averaging over their individuals properties.

Previous works that computed average rest-frame FUV-to-FIR LBG SEDs (e.g., \citealt{Magdis2010c}) performed a simple average to the optical/NIR photometry. Our LBG sample is 70-80\% complete at NIR and 90-100\% complete at optical bands. Then, we decided to perform a stacking analysis in the optical/NIR to eliminate a possible bias to the brightest LBGs (if only detected LBGs are averaged).

We split our LBG sample as a function of $L_{\mathrm{FUV}}$, $\beta_{\mathrm{UV}}$, and M$_{*}$ in five different ways. The first three are done as a function of the single parameters $L_{\mathrm{FUV}}$, $\beta_{\mathrm{UV}}$, and M$_{*}$. For the two others, we split the sample as a function of the combination of  $\beta_{\mathrm{UV}}$ and M$_{*}$ in the ($\beta_{\mathrm{UV}}$, M$_{*}$) plane. Table \ref{tabla_sample_bining_5} shows the sample binning. It includes the interval used to define each bin, the number of LBGs that are stacked, and the mean values of $L_{\mathrm{FUV}}$, $\beta_{\mathrm{UV}}$, M$_{*}$, and redshift that define each bin. The uncertainties associated with the average values are defined by the standard deviation of the mean. The number of bins and the intervals are optimized to obtain a good signal-to-noise ratio (S/N) on the final FIR (SPIRE bands) stacked images.

\begin{table*}[h]
\caption{\label{tabla_sample_bining_5}Sample binning}
\centering
\resizebox{0.75\textwidth}{!}{
\begin{tabular}{ccccccc}
\hline
ID & Range & $\log$($L_{\mathrm{FUV}}$[L$_{\odot}$]) & $\log$($M_{*}$ [M$_{\odot}$]) & $\beta_{\mathrm{UV}}$ & $z$ & N$_{gal}$ \\
\hline
\multicolumn{7}{c}{Stacking as a function of $L_{\mathrm{FUV}}$ (LBG-$L$)}\\
\hline

LBG-$L$1 & 10.2 - 10.5 & 10.35$\pm$0.09 & 9.67$\pm$0.47 & -1.51$\pm$0.45 & 3.00$\pm$0.23 & 9145  \\
LBG-$L$2 & 10.5 - 10.8 & 10.63$\pm$0.09 & 9.78$\pm$0.44 & -1.56$\pm$0.41 & 3.08$\pm$0.25 & 5385 \\
LBG-$L$3 & 10.8 - 11.1 & 10.91$\pm$0.08 & 9.94$\pm$0.37 & -1.54$\pm$0.36 & 3.11$\pm$0.25 & 1510 \\
LBG-$L$4 & 11.1 - 11.4 & 11.20$\pm$0.08 & 10.14$\pm$0.38  & -1.54$\pm$0.32 & 3.15$\pm$0.24 & 199  \\
\hline
\multicolumn{7}{c}{Stacking as a function of $\beta_{\mathrm{UV}}$ (LBG-$\beta$)}\\
\hline
LBG-$\beta$1 & -1.7 - -1.1 & 10.49$\pm$0.24 & 9.75$\pm$0.44 & -1.42$\pm$0.17 & 2.99$\pm$0.25 & 8531  \\ 
LBG-$\beta$2 & -1.1 - -0.7 & 10.41$\pm$0.22 & 9.88$\pm$0.62 & -0.94$\pm$0.11 & 2.94$\pm$0.24 & 2445  \\
LBG-$\beta$3 & -0.7 - -0.3 & 10.30$\pm$0.20 & 10.00$\pm$0.66 & -0.54$\pm$0.10 & 2.91$\pm$0.21 & 593 \\
LBG-$\beta$4 & -0.3 -  0.3 & 10.26$\pm$0.21 & 10.17$\pm$0.72 & -0.10$\pm$0.21 & 2.89$\pm$0.21 & 114  \\
\hline
\multicolumn{7}{c}{Stacking as a function of stellar mass (LBG-$M$)}\\
\hline
LBG-$M$1 & 9.75-10.00 &  10.51$\pm$0.24 &  9.86$\pm$0.08 & -1.45$\pm$0.38 & 3.05$\pm$0.25 & 4517 \\
LBG-$M$2 & 10.00-10.25 & 10.55$\pm$0.26 & 10.10$\pm$0.08 & -1.29$\pm$0.41 & 3.02$\pm$0.26 & 2257 \\
LBG-$M$3 & 10.25-10.50 & 10.58$\pm$0.29 & 10.35$\pm$0.08 & -1.14$\pm$0.43 & 2.98$\pm$0.26 & 1041 \\
LBG-$M$4 & 10.50-10.75 & 10.55$\pm$0.27 & 10.60$\pm$0.08 & -1.08$\pm$0.45 & 2.95$\pm$0.26 &  372 \\
LBG-$M$5 & 10.75-11.00 & 10.52$\pm$0.31 & 10.86$\pm$0.08 & -0.95$\pm$0.43 & 2.94$\pm$0.25 &  160 \\
LBG-$M$6 & 11.00-11.50 & 10.47$\pm$0.28 & 11.16$\pm$0.12 & -0.77$\pm$0.51 & 2.96$\pm$0.24 &   55 \\
\hline
\multicolumn{7}{c}{Stacking as a function of ($\beta_{\mathrm{UV}}$, M$_{*}$) - LBG-$M\beta$1}\\
\hline
LBG-$M\beta$1-$M_0 \beta_0$ & (-1.7 - -1, 9.75-10.65) &  10.55$\pm$0.25 &  10.00$\pm$0.20 & -1.35$\pm$0.20 & 3.02$\pm$0.25 & 4799 \\
LBG-$M\beta$1-$M_0 \beta_1$ & (-1 - -0.5, 9.75-10.65) &  10.41$\pm$0.22 &  10.12$\pm$0.23 & -0.82$\pm$0.14 & 2.93$\pm$0.24 & 1343 \\
LBG-$M\beta$1-$M_0 \beta_2$ & (-0.5 - 0.3, 9.75-10.65) & 10.26$\pm$0.19 & 10.18$\pm$0.23 & -0.32$\pm$0.16 & 2.87$\pm$0.21 & 208 \\
LBG-$M\beta$1-$M_1 \beta_0$ & (-1.7 - -1, 10.65-11.50) &  10.62$\pm$0.31 &  10.83$\pm$0.16 & -1.28$\pm$0.19 & 2.97$\pm$0.26 & 146\\
LBG-$M\beta$1-$M_1 \beta_1$ & (-1 - -0.5, 10.65-11.50) &  10.44$\pm$0.20 &  10.84$\pm$0.16 & -0.78$\pm$0.14 & 2.89$\pm$0.23 & 122\\
LBG-$M\beta$1-$M_1 \beta_2$ & (-0.5 - 0.3, 10.65-11.50) &  10.35$\pm$0.20 &  10.95$\pm$0.20 & -0.41$\pm$0.05 & 2.90$\pm$0.22 & 40 \\
\hline
\multicolumn{7}{c}{Stacking as a function of ($\beta_{\mathrm{UV}}$, M$_{*}$) - LBG-$M\beta$2}\\
\hline
LBG-$M\beta$2-$M_0 \beta_0$ & (-1.7 - -0.8, 9.75-10.125) &  10.51$\pm$0.23 &  9.91$\pm$0.11 & -1.31$\pm$0.23 & 3.01$\pm$0.25 & 4056 \\
LBG-$M\beta$2-$M_0 \beta_1$ & (-0.8 - 0.3, 9.75-10.125) &  10.27$\pm$0.16 &  9.95$\pm$0.11 & -0.60$\pm$0.19 & 2.94$\pm$0.22 &  338 \\
LBG-$M\beta$2-$M_1 \beta_0$ & (-1.7 - -0.8, 10.125-10.5) &  10.61$\pm$0.28 & 10.27 $\pm$0.10 & -1.22$\pm$0.23 & 2.99$\pm$0.26 &  1373\\
LBG-$M\beta$2-$M_1 \beta_1$ & (-0.8 - 0.3, 10.125-10.5) &  10.37$\pm$0.21 &  10.29$\pm$0.10 & -0.56$\pm$0.21 & 2.88$\pm$0.21 &  337 \\
LBG-$M\beta$2-$M_2 \beta_0$ & (-1.7 - -0.8, 10.5-10.75) &  10.59$\pm$0.27 & 10.61$\pm$0.07 &   -1.19$\pm$0.25 & 2.96$\pm$0.26 & 260 \\
LBG-$M\beta$2-$M_2 \beta_1$ & (-0.8 - 0.3, 10.5-10.75) &  10.40$\pm$0.22 &  10.60$\pm$0.08 & -0.52$\pm$0.23 & 2.87$\pm$0.25 &  86 \\
LBG-$M\beta$2-$M_3 \beta_0$ & (-1.7 - -0.8, 10.75-11) &  10.59$\pm$0.31 &  10.85$\pm$0.07 & -1.18$\pm$0.23 & 2.94$\pm$0.25 &  102 \\
LBG-$M\beta$2-$M_3 \beta_1$ & (-0.8 - 0.3, 10.75-11) &  10.35$\pm$0.18 &  10.86$\pm$0.08 & -0.49$\pm$0.27 & 2.91$\pm$0.23 & 53 \\
LBG-$M\beta$2-$M_4 \beta_0$ & (-1.7 - -0.8, 11-11.5) &  10.58$\pm$0.29 &  11.14$\pm$0.13 & -1.15$\pm$0.24 & 3.04$\pm$0.26 &  27 \\
LBG-$M\beta$2-$M_4 \beta_1$ & (-0.8 - 0.3, 11-11.5) &  10.40$\pm$0.24 &  11.18$\pm$0.12 & -0.38$\pm$0.26 & 2.88$\pm$0.18 & 26 \\
\hline
\end{tabular}}

 \label{TabLFs}
\end{table*}

\subsection{General method}\label{general_stacking}

The general method used to perform the optical-to-FIR stacking analysis is based on the IAS library (Bavouzet 2008 and \citealt{Bethermin2010b})\footnote{\url{http://www.ias.u-psud.fr/irgalaxies/downloads.php}}. Given a specific catalog and field image, the library extracts square cutout images centered on each source and stores these images in a cube. Then, it generates a stacked image by averaging them. Additionally, it rotates each image by +$\pi$/2 with respect to the previous image to cancel out the large-scale background gradients. To get valid and reliable results, the general method has to be corrected for different effects that generate a nonhomogeneous background on the stacked image (Bavouzet 2008; \citealt{Bethermin2010b}; \citealt{Heinis2013}; AM16). The following three points are of major importance: 

\begin{enumerate}
\item Correction for clustering of the input catalog. The LBGs are clustered between them and other star-forming galaxies at high redshift. Depending on the PSF size of the available observations, the PSF beam may be contaminated by different nearby sources included in the same sample. Therefore, we could account for the flux of the same source more than once during the stacking analysis. This effect would lead to an overestimation of the mean flux due to the clustered nature of the sources. It is corrected by taking into account the angular correlation function of the input catalog. The complete formalism is detailed in AM16.
\item Correction for incompleteness of the input catalog in the dense regions. A bias is produced in the stacked image when the population of sources is not complete (\citealt{Dole2006}; Bavouzet 2008; \citealt{Bethermin2010b}; \citealt{Heinis2013}; \citealt{Viero2014}). If the stacked sample presents a relatively high percentage of incompleteness, part of the faint LBG population (mostly located in the crowded areas of the image or close to bright objects) are missed during the source extraction process at optical images. When we move to longer wavelength observation (e.g., FIR), the optical crowded areas and the bright objects contribute to enlarge the local background level. This means that if we perform the stacking analysis on an incomplete population, we would mostly stack the areas in which the background presents lower levels and lose the contribution of the areas with higher levels. It produces a negative contribution in the central part of the stacked image related to the global background level. We corrected this effect using extensive simulations to characterize the detection process and sample selection. We performed these simulations by injecting mock sources in the original images ($V_{\mathrm J}$ and $i^{+}$ bands), and keeping track of the recovered sources. The full formalism used to correct this effect is deeply explained in AM16.

\item Correction for bright field sources in crowded field observations. A deep observation, not dominated by the confusion noise, generates a crowded field image. The presence of a large number of bright neighbors may generate an additional noise level that perturbs the final stacked image. We removed the contribution of field sources in our stacking procedure by performing a weighted stacking analysis to mask them. We proceeded as follows:

\begin{itemize}
\item We generated a segmentation map, using SExtractor \citep{BertinArnouts1996}, on the image where the stacking analysis is performed. A source extraction was applied to detect all the objects with flux larger than 3$\sigma_{Back}$, excluding the LBGs from our sample. 
\item We created a weight map by assigning a value equal to 1 for background pixels and equal to 0 for detected source pixels.
\item We performed the stacking analysis by using a weighted mean, $\sum_{i}I_i W_i / \sum_{i}W_i$, for each individual pixel of the cutout image. Where $i$ corresponds to each object inside the sample, $I_i$ one of the pixels from the cutout image extracted from the map, and $W_i$ one of the pixels from the cutout image extracted from the weighted map. 
\end{itemize}

  \begin{figure}[h]
   \centering

   \includegraphics[width=\hsize]{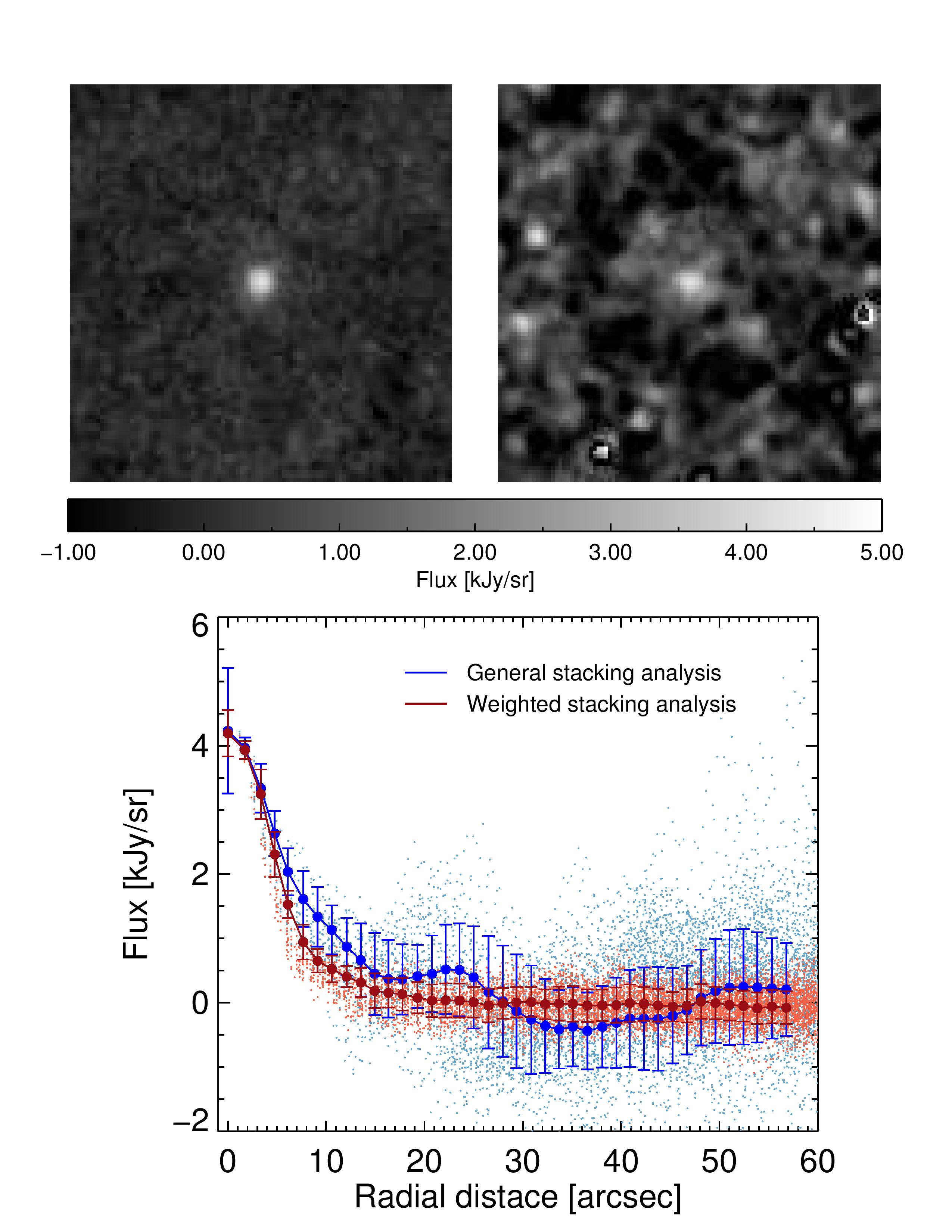}
      \caption{Top panel: Maps of the final MIPS stacked images for the first bin as a function of the $L_{\mathrm{FUV}}$ ($10.2 < log(L_{\mathrm{FUV}}[L_{\odot}]) > 10.5$). The left panel shows the stacked image using the weighted stacking analysis and the right image using the general stacking method. The bottom panel shows the radial profiles estimated from the upper images. The blue line corresponds to the stacking analysis performed using the general method, and the red line correspond to the stacking analysis performed using the weighted method. The lighter blue and red points correspond to the value of each pixel from the stacked images.} 
         \label{mips_profiles}
   \end{figure}

Figure \ref{mips_profiles} shows the improvement that we get in the final stacked image using a weighted stacking analysis for the 24 $\mu$m MIPS observations. This example corresponds to the final stacked image, with and without applying the weighted procedure, for one of the bins of the stacking analysis as a function of L$_{\mathrm{FUV}}$. This new method provides a considerable improvement in the background noise, which is essential to detect and recover the emission of the stacked population.
\end{enumerate}

Each of these corrections impacts at various levels the final stacked image, which mainly depends on the image quality (PSF, confusion noise, and depth), observed wavelength, and sample selection method. In Sections \ref{IRstacking}, \ref{midIRstacking}, and \ref{opticalstacking}, we discuss which of the previous corrections are taken into account during the stacking analysis for each of the data sets.

\subsection{Far-infrared analysis}\label{IRstacking}

The following procedure is applied to perform the stacking analysis on PACS (100 and 160 $\mu$m) and SPIRE (250, 350, and 500 $\mu$m) images. It was previously used and explained well in AM16. We used the calibrated PACS and SPIRE maps without cleaning the images from any detected source, and we did not apply the correction for bright field sources in crowded field observations to mask the bright neighbors during the stacking procedure. If a large number of sources were stacked in confusion limit maps, as in our FIR images, the bright neighbors tend to average homogeneously to a constant background level. We demonstrated this by performing a random stacking analysis at the same number of positions than objects in each of our sample bins.  

The PACS and SPIRE observations present large PSF sizes, with FWHM PSFs that vary from 6$^{\prime\prime}$ to 36.3$^{\prime\prime}$. Our LBG sample is clustered and could be defined by an autocorrelation function, $\omega [\theta,\phi] \propto \theta^{-\gamma}$, with $\gamma$ = 0.63 (AM16). The combination of large PSF sizes and the clustered nature of our LBG sample could produce a contamination of our LBG PSF beams due to a nearby LBGs. If we stack them, we account for the flux of the same source more than once during the stacking analysis. Therefore, we applied the correction for clustering of the input catalog to deconvolve the emission of our LBG population and the clustering contribution. Additionally, we corrected our stacking analysis for the incompleteness of the input catalog in the dense regions following the analysis done in AM16.

\subsection{Mid-Infrared analysis}\label{midIRstacking}

The following procedure is applied to perform the stacking analysis for IRAC (3.6, 4.5, 5.8, 8 $\mu$m) and MIPS (24 $\mu$m) observations. The IRAC images present a PSF FWHM equal to 1.66$^{\prime\prime}$, 1.72$^{\prime\prime}$, 1.88$^{\prime\prime}$, and 1.98$^{\prime\prime}$ with a pixel size of 0.6$^{\prime\prime}$, and the MIPS observation has a PSF FWHM equal to 6$^{\prime\prime}$ and a pixel size of 1.2$^{\prime\prime}$. The improvement of the PSF FWHM size and sensitivity from FIR to MIR observations reduces the confusion noise and improves the quality of the MIR images. This generates a crowded image with a large number of field sources. The presence of the large amount of field sources in the maps generates an additional noise level and clustering contribution that perturbs our stacked images. To remove this effect from our stacked sources, we applied the weighted stacking analysis previously presented in the correction for bright field sources in crowded field observations.

The reduction of the PSF FWHM sizes from FIR to MIR maps lowers and makes negligible the clustering effect on the stacked LBGs. Therefore, here, we did not take into account any corrections related to the clustering nature from our input LBGs population in the stacked LBGs. However, the bias induced in our stacked image due to the incompleteness of our input catalog is still present. We realized that this effect decreases when the stacking procedure is performed closer to the sample selection wavelengths ($V_{\mathrm J}$ and $i^{+}$ bands). However, it is still important to correct for this effect in the MIR wavelength.

We performed an aperture photometry to calculate the flux density of the stacked LBGs in the IRAC and MIPS observations. We made a circular aperture with radius equal to 3.6$^{\prime\prime}$ and 7$^{\prime\prime}$, and a background annulus from 3.6$^{\prime\prime}$ to 8.4 $^{\prime\prime}$ and 20$^{\prime\prime}$ to  32$^{\prime\prime}$. We then applied an aperture correction of  1.125, 1.120, 1.135, 1.221, and 1.61 at IRAC (3.6, 4.5, 5.8, 8 $\mu$m) and MIPS (24 $\mu$m)\footnote{The photometry configuration and aperture correction have been obtained from the instrument handbook of IRAC (\url{http://irsa.ipac.caltech.edu/data/SPITZER/docs/irac/iracinstrumenthandbook/27/}), and MIPS (\url{http://irsa.ipac.caltech.edu/data/SPITZER/docs/mips/mipsinstrumenthandbook/50/})}. We used bootstrap resampling to obtain the mean values and errors. We repeated the above procedure over 3000 random bootstrap samples and adopted 1$\sigma$ of the distribution of the derived fluxes as the uncertainty for our results. Appendix \ref{table_stak_ir} presents the derived fluxes and uncertainties of our stacking analysis at FIR and MIR. This appendix includes the stacking analysis performed for AzTEC (1.1mm), and previously presented in AM16.

\subsection{Optical/near-infrared analysis}\label{opticalstacking}

The following procedure is applied to perform the stacking analysis in the optical (B$_{\mathrm J}$, V$_{\mathrm J}$, r$^{+}$, i$^{+}$, z$^{++}$ bands) and NIR (Y, J, H, K$_{s}$) observations. The optical and NIR images have a homogenized PSF with a FWHM equal to 0.8$^{\prime\prime}$, and a large depth improvement with respect to the MIR observations. The optical/NIR observations contain large number of bright field sources that are removed by performing a weighted stacking analysis. We did not apply any corrections related to the clustering nature of our sample and the incompleteness of our input catalog. These two effects are negligible in our stacked optical/NIR images.

We computed the photometry on the stacked object by fitting a Moffat function. The error is calculated by combining the contribution of different effects: bootstrap, error in the fitting procedure, and Poisson noise. Appendix \ref{table_stak_optical} presents the derived fluxes and uncertainties of our stacking analysis in optical and NIR wavelengths. We verified the method by comparing the derived stacked fluxes with an average observed photometry of individuals LBGs at $V_{\mathrm J}$ and $i^{+}$ bands, where the sample is complete. In both cases, we obtain results within the derived uncertainties. 

\section{SED-fitting analysis}\label{sed_fitting}

A Python-based modular code, CIGALE \citep{Burgarella2005, Noll2009, Boquien2019}\footnote{Code Investigating
GALaxy Emission (CIGALE): \url{https://cigale.lam.fr/}} is  dedicated to fitting SEDs from the UV to the FIR and to creating galaxy emission models in the same wavelength range. The CIGALE code is  modular and allows the user to create models by selecting independent modules for parameters such as the SFH, stellar models, shape of the dust attenuation curve, nebular emission, contribution of an active galactic nuclei, and dust emission templates or models.

\subsection{Initial parameters }\label{initial_param_sedfitting}

Deriving physical properties of galaxies by fitting synthetic SEDs to a multiband photometry requires making prior assumptions on the initial parameters to build the synthetic SED library. Assumptions such as SFH, IMF, metallicity, and dust attenuation curve affect the derived galaxy properties like the SFR, stellar mass, age of the stellar population, amount of dust attenuation (e.g., \citealt{Papovich2001, Maraston2010, PforrMaraston2012, Schaerer2013,Buat2014,LoFaro2017}). Table \ref{Ini_param} summarizes the initial parameters used to perform the SED-fitting analysis for our stacked LBGs SEDs at $z\sim3$.

\begin{table}[h]
\caption{\label{Ini_param}Initial parameters used in the SED-fitting analysis}
\centering
\renewcommand{\arraystretch}{0.8}
\begin{tabular}{lcl}
\hline
Parameter & Symbol & Range \\
\hline
\multicolumn{3}{c}{SFH (Delay-$\tau$)}\\
\hline
age & age & 10 to 2000 Myr \\
e-folding timescales & $\tau$ & 20 to 1500 Myr\\
\hline
\multicolumn{3}{c}{SSP \citep{BruzualCharlot2003}}\\
\hline
Initial mass function & IMF & \cite{Chabrier2001}\\
Metallicity & Z & Fixed to 0.2Z$_{\odot}$\\
\hline
\multicolumn{3}{c}{Dust attenuation \citep{Noll2009}}\\
\hline
Color excess  & E(B-V) & 0.025 to 1.5 \\
Slope of the power law & $\delta$ & -0.5 to 0.5\\
UV bump at 2175$\AA$ & & Not included \\
\hline
\multicolumn{3}{c}{Dust emission \citep{DraineLi2007} }\\
\hline
Mass fraction of PAH & p$_{PAH}$ & fixed to 3.9\% \\
Minimum radiation field & U$_{min}$ & 0.7 to 50 \\
Power slope $dU/dM\propto U^{-\alpha}$ & $\alpha$ & fixed to 2.0 \\
Dust fraction in PDRs & $\gamma$ & 0.01 to 0.04 \\
\hline
\end{tabular}
\end{table}

We adopted the stellar population models from \cite{BruzualCharlot2003} and an IMF from \cite{Chabrier2001}. Delayed SFHs (SFR $\propto t/\tau^{2} ~e^{-t/\tau}$), with varying e-folding times are assumed to model the SFH of our LBGs at $z\sim3$ \citep{Lee2010}. We performed a test to check the influence of the chosen SFH on the final derived physical parameters, where we ran CIGALE with different SFHs recipes (exponential declining, exponential rising, constant, and delayed) using the same initial parameters from Table \ref{Ini_param}. The results suggest that our SED-fitting analysis cannot distinguish which of the tested SFHs recipes are more likely to reproduce the LBG population. Nevertheless, our SED-fitting analysis can derive fully consistent physical parameters independent of the SFH chosen with the exception of the age of the stellar population. 

The metallicity was fixed to 0.2 Z$_{\odot}$ \citep{Castellano2014}. We fixed the metallicity because i) the SED-fitting technique applied to broadband photometry can hardly constrain the metallicity due to the age-dust-metallicity degeneracy \citep{Lopez-Fernandez2016}, and ii) the metallicity of our stacked SEDs are averaged due to the combination of a large number of galaxies. However, there is a possibility to have an evolution of metallicity in our LBG sample due to the well-known mass-metallicity relation (e.g., \citealt{Mannucci2009}). Then, we performed an individual SED-fitting analysis for the stacking analysis as a function of M$_{*}$ with the metallicity as a free parameter. The derived physical parameters are in agreement, within the uncertainties, with those from the fixed metallicity. Therefore, we concluded that our choice of fixed metallicity does not have a large impact on the final derived physical parameters. 

  \begin{figure*}[h]
  \centering
  \includegraphics[width=\hsize]{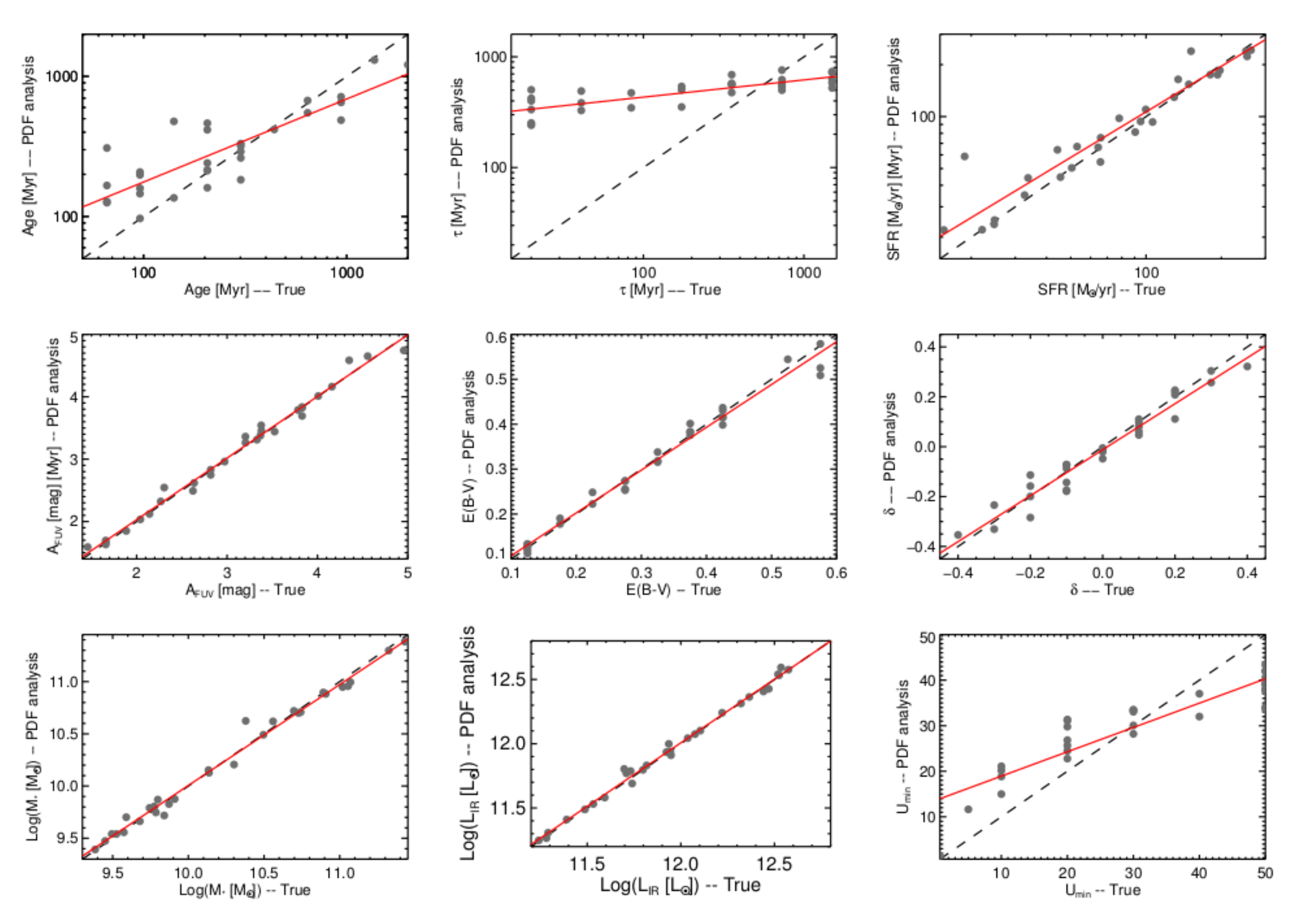}
   \caption{CIGALE mock analysis results. On the x-axis the \textit{true} parameter values are provided by the best-fit model + noise consitute the mock SED catalog. These true values are compared with the PDF value estimated by CIGALE on the y-axis. The 1-1 correlation line is shown as a black long dashed line in each panel. From top left to bottom right, we show the results for the age of the stellar population, e-folding timescale of delay-$\tau$ SFH, SFR, FUV dust attenuation, E(B-V), slope of the dust attenuation curve - $\delta$, stellar mass, total IR luminosity, and U$_{min}$. The regression lines for each assumed configuration are also plotted as red solid lines.} 
  \label{comparison_SEDfitting_parameters_Delay}
   \end{figure*}

Dust attenuation is treated using the prescription given by \cite{Noll2009} in Eq. \ref{noll_dust_att_eq}. These authors used the Calzetti dust attenuation law (k($\lambda$), \citealt{Calzetti2000}) as a baseline. This provides the possibility to i) include a UV bump at 2175 $\AA$ modeled by a Lorentzian-like profile, and ii) modify the slope of the dust attenuation curve by a power law,

\begin{equation}\label{noll_dust_att_eq}
A (\lambda) = \frac{A_{V}}{4.05} \left( k(\lambda) + D_{\lambda_{o}, \gamma, E_{bump}} \right) \left(  \frac{\lambda}{\lambda_{V}} \right)^{\delta}
,\end{equation} 

where $A(\lambda)$ is the modified dust attenuation curve presented by \cite{Noll2009}, $A_{V}$ is the dust attenuation in the V band, $D_{\lambda_{o}, \gamma, E_{bump}}$ is the UV bump profile, $\delta$ is the power-law slope with respect to the Calzetti's law, and $\lambda_{V}$ = 5500$\AA$. The shape of the dust attenuation curve depends on the stardust geometry, grain size distribution, etc. \citep{Witt2000}. Complex dust geometries can produce a grayer or shallower dust attenuation curve than Calzetti's law \citep{Buat2011a, KriekConroy2013, Zeimann2015,Salmon2016,LoFaro2017}. Our SED fitting takes $\delta$ as a free parameter to mimic the variation on the shape of the dust attenuation curve. The absorption feature produced by dust at 2175 $\AA$, UV bump, is set to zero because the stacked SEDs do not have enough spectral resolution to constrain it. 

The IR emission is modeled using \cite{DraineLi2007} dust models. The validity of these models to reproduce the FIR emission of Herschel detected high-z main sequence galaxies was confirmed by \cite{Magdis2012}. The dust models require the fine-tuning of several free parameters: the mass fraction of PAH (polycyclic aromatic hydrocarbons, q$_{PAH}$), minimum and maximum radiation field (U$_{min}$, U$_{max}$), power slope $dU/dM\propto U^{-\alpha}$ ($\alpha$), dust fraction in PDRs ($\gamma$), and dust mass (M$_{dust}$). However, \cite{DraineLi2007} showed that fixed values of $\alpha = 2$ and $U_{max} = 10^{6}$ can reproduce the SEDs of galaxies with a wide range of properties. The stacked SEDs only have the information of MIPS 24 $\mu$m (rest-frame~6$\mu$m) in the PAH part of the spectrum, which is not enough to constrain the q$_{PAH}$. After some initial tests with the q$_{PAH}$ as a free parameter, we fixed q$_{PAH}$ to 3.9\%. This value best fit our SEDs by eye even though it is higher than that proposed by \cite{Magdis2012}, who found an interval from 1.12\% to 3.19\%. We consider U$_{min}$ to vary between 0.7 to 50. \cite{DraineLi2007} suggested using $U_{min} > 0.7$ because lower values correspond to dust temperatures below $\sim$ 15K, which we cannot constrain by our FIR photometry alone. The fraction of dust enclosed in the PDRs was fixed by \cite{Magdis2012} to $\gamma$=0.02, and \cite{DraineLi2007} suggested that the $\gamma$ value should not be higher than 0.04 (4\%). We detected some variations at rest-frame wavelength 20-50 $\mu$m for our set of stacked SEDs, which suggest different $\gamma$ values. Therefore, we used a $\gamma$ range from 0.01 to 0.04. 

The adopted initial configuration (Table \ref{Ini_param}) gives a total of 609840 synthetic galaxy models/SEDs. This set of models is used to perform the SED fitting and Bayesian analysis over our 30 rest-frame FUV-to-FIR stacked LBGs SEDs at $z\sim3$.
  
\subsection{Physical parameters for LBGs at $z\sim3$} \label{physical_param_sed_fitting}

We ran CIGALE for the sample of 30 rest-frame FUV-to-FIR stacked LBGs SEDs at $z\sim3$. We used the adopted initial configuration listed in Table \ref{Ini_param}. The CIGALE code estimates and derives the physical parameters, such as stellar mass, dust mass, age of stellar populations, SFR, rest-frame IR and FUV luminosities, FUV attenuation, and $\beta_{\mathrm{UV}}$, as well as the input parameters from Table \ref{Ini_param}. Additionally, CIGALE provides the synthetic model that best fits the observational SEDs by $\chi^{2}$ minimization.

The CIGALE code uses a Bayesian analysis to derive the physical parameters. This code builds the probability distribution function (PDF) for each parameter, and takes its mean and standard deviation \citep{Noll2009,Boquien2019}. We used a mock analysis to check the robustness, accuracy, and parameter degeneracy of our estimates. The mock analysis consists in generating a mock catalog of artificial SEDs, similar to the input LBGs, for which the physical parameters have been previously derived, and adding a Gaussian noise where the dispersion is taken from the observed uncertainty. Then, we compared these values with the CIGALE estimates (more details: \citealt{Buat2012, Ciesla2015,LoFaro2017,Boquien2019}). This procedure allows us to check our ability in estimating and constraining the individual output parameters given the information provided by the detailed shape of the broadband SED. The results are summarized in Figure \ref{comparison_SEDfitting_parameters_Delay}. We call the parameter value provided by the best-fit model ``true'' and used this value to compute the mock SEDs. This ``true'' parameter value, on the x-axis, is compared with the PDF value, on the y-axis, which is computed by SED-fitting analysis over the mock SEDs catalog. From top left to bottom right, Figure \ref{comparison_SEDfitting_parameters_Delay} shows the results for the age of the stellar population, e-folding timescale of delay-$\tau$ SFH, SFR, FUV dust attenuation, E(B-V), change of the slope of the dust attenuation law with respect to Calzetti - $\delta$, stellar mass, total IR luminosity, and U$_{min}$. On the one hand, the results emphasize that we are not able to determine the shape of the SFH and age of the stellar population by SED-fitting analysis. The estimated PDFs of the e-folding timescale, $\tau$, appear to average out to a constant value independent of the true value used to compute the mock SEDs. The age of the stellar population is slightly constrained for the particular case of delay-$\tau$ SFH model, but it is biased against the chosen SFH. On the other hand, the results suggest that our procedure well constrain the physical parameters: stellar mass, IR luminosity, FUV dust attenuation, SFR, and change of the slope of the dust attenuation law with respect to Calzetti  - $\delta$, with the chosen delay-$\tau$ SFH. In Section \ref{initial_param_sedfitting}, we also conclude that these physical parameters are not affected by the chosen SFH. Therefore, despite of the degeneracy for the SFH and ages of the stellar population, our SED-fitting analysis provides consistent physical parameters. 

Appendix \ref{physical_param} lists the input and additional physical parameters derived by SED-fitting analysis in the 30 stacked LBGs SEDs. The $\beta_{\mathrm{UV}}$ and L$_{\mathrm{UV}}$ are computed in a separate FUV-optical SED-fitting analysis; the input parameters and procedure are already explained in Section \ref{slope_uvluminosity}. 

\subsection{Templates of LBGs at $z\sim3$}\label{templates_LBGs}

We used the results from the panchromatic stacking and SED-fitting analysis to build a library of empirical rest-frame FUV-to-FIR templates of LBGs at $z\sim3$.\footnote{Templates  and  SEDs  are  only  available  at  the  CDS via  anonymous  ftp  to \url{cdsarc.u-strasbg.fr} ($130.79.128.5$)  or  via  \url{http://cdsarc.u-strasbg.fr/viz-bin/cat/J/A+A/630/A153}} The library contains 30 empirical SEDs and their best-fit synthetic model spectrum. The empirical SEDs correspond to the rest-frame FUV-to-FIR stacked LBGs SEDs derived by stacking analysis in Section \ref{stacking_analysis}. These SEDs result from combining $\sim$17000 individual LBGs in different bins as a function of L$_{\mathrm{FUV}}$, M$_{*}$, $\beta_{\mathrm{UV}}$, and a combination of M$_{*}$ and $\beta_{\mathrm{UV}}$. More specifically, the number of individuals LBGs used to obtain each stacked LBG SED varies from about 9000 to 100. This statistics strengthens our library in the sense that our SEDs have an exceptional S/N, which is not comparable with any individual detected LBG at $z\sim3$. The set of synthetic model spectrum associated with each stacked LBG SED is obtained with CIGALE.

Thanks to the binning configuration used to split our large LBG sample, we derived a set of templates with a large variety of physical properties. Our library contains templates within an interval of stellar mass, $9.2 < \log(M_{*}\,[M_{\odot}]) < 11.4$; SFR, $20 < SFR\, [M_{\odot}yr^{-1}] < 300$,  $\beta_{\mathrm{UV}}$,  $ -1.8 < \beta_{\mathrm{UV}} < -0.2$; FUV dust attenuation, $1.5 < A_{\mathrm{FUV}}\, [mag] < 4.8$; IR luminosity, $11.2 <  \log(L_{\mathrm{IR}} \,[L_{\odot}]) < 12.7$; and FUV luminosity, $10.4 <  \log(L_{\mathrm{FUV}} \,[L_{\odot}]) < 11.2$. This variety makes our library versatile in the sense that our templates fit a wide range of physical properties for star-forming galaxies at $z\sim3$.

\begin{figure}[h]
   \centering
   \includegraphics[width=\hsize]{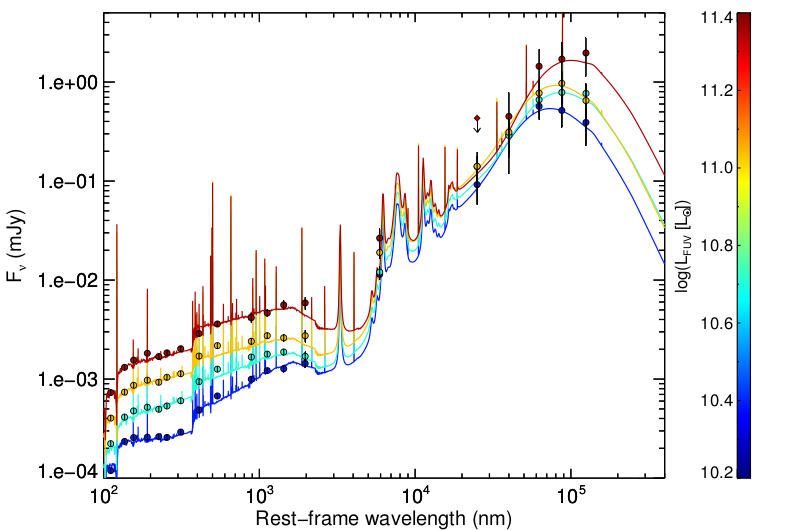}
      \caption{Stacked LBGs SEDs and the best-fit models for the stacking as a function of L$_{\mathrm{UV}}$. } 
         \label{sed_uvlum_model}
  \end{figure}

Figure \ref{sed_uvlum_model} shows the four SEDs and templates derived in the stacking analysis as a function of $L_{\mathrm{FUV}}$. If a sample binning is carried out as a function of $L_{\mathrm{FUV}}$, the subsample becomes a mix of red, blue, high, and low-mass galaxies. For example, a massive red galaxy with a large dust attenuation may have the same $L_{\mathrm{FUV}}$ than a low-mass blue galaxy. Therefore, if a stacking analysis is applied to a mix population of LBGs, the final results tend to average to a similar mean SED with a different normalization factor, i.e., $L_{\mathrm{FUV}}$.  Figure \ref{sed_uvlum_model} shows the rest-frame FUV-to-NIR part of the SEDs presents a similar behavior except for the lowest $L_{\mathrm{FUV}}$ bin, which has a slightly bluer $\beta_{\mathrm{UV}}$. We can also see this from the derived physical properties on the SED-fitting analysis, where we only find differences  in the bolometric luminosity, M$_{*}$ and SFR, which is associated with an increase of the normalization factor. However, the FIR peak has progressively shifted to longer wavelengths for higher $L_{\mathrm{FUV}}$, which suggests a lower mean dust temperature. If we consider that the U$_{min}$ values are correlated with the location of the FIR peak \citep{DraineLi2007, Magdis2012}. This behavior is reinforced by the variation of the U$_{min}$ values (38.5 < U$_{min}$ < 19.2), derived from our SED-fitting analysis, for the lowest to the highest $L_{\mathrm{FUV}}$ bin.

      \begin{figure*}[h]
   \centering
   \includegraphics[width=\hsize]{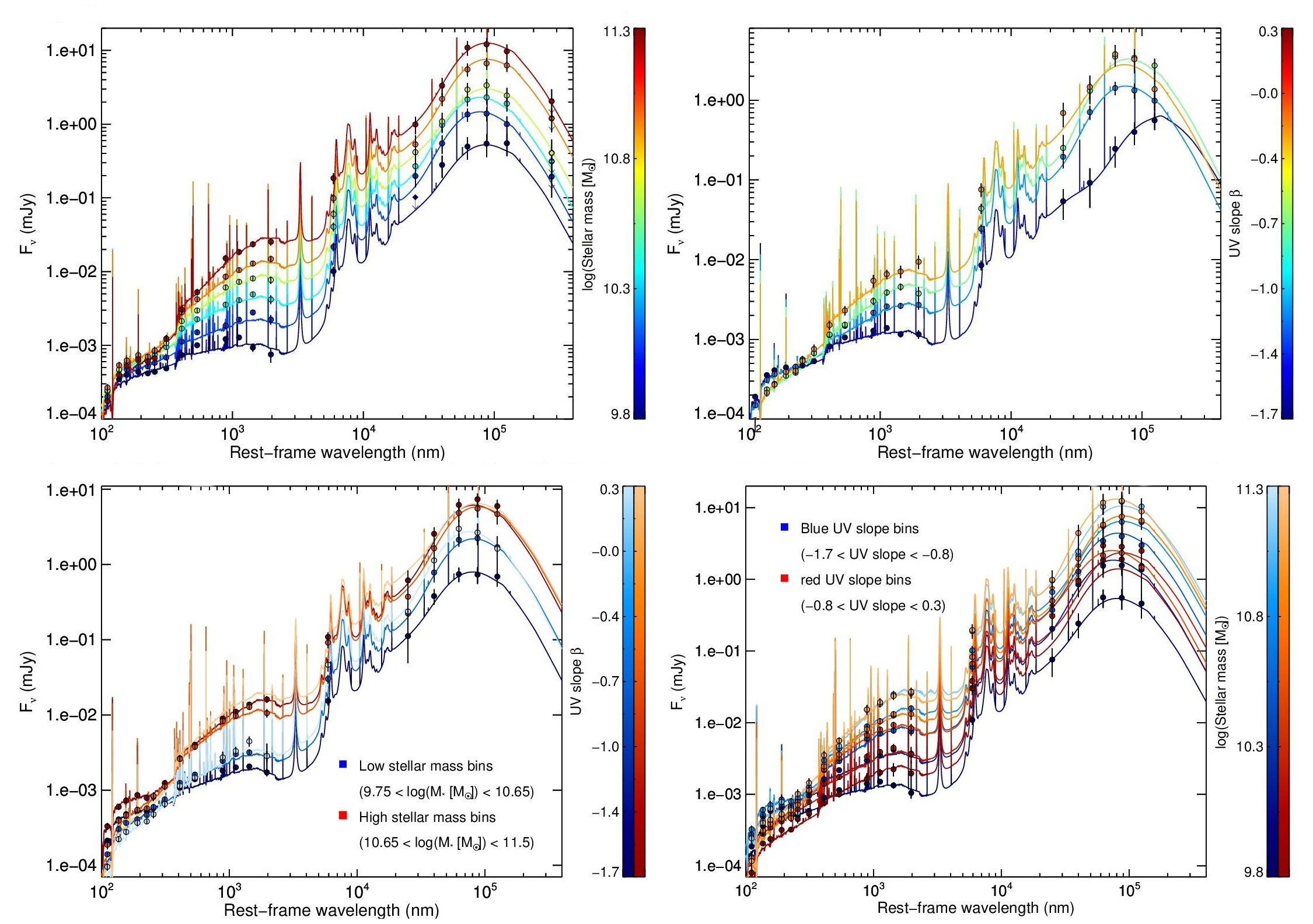}
      \caption{Stacked LBGs SEDs and the best-fit models for the stacking analyses as a function of M$_{*}$ (top left  panel) and  $\beta_{\mathrm{UV}}$ (top right panel). The bottom panel shows the LBG-$M\beta$1 (left) and LBG-$M\beta$2 (right) stacking analyses, which are derived from a combination of both, M$_{*}$ and  $\beta_{\mathrm{UV}}$, in the M$_{*}$ and  $\beta_{\mathrm{UV}}$ plane.} 
         \label{sed_mass_beta_model}
  \end{figure*}
  
The top left panel of Figure \ref{sed_mass_beta_model} illustrates the six SEDs and templates derived from the stacking analysis as a function of M$_{*}$. Their rest-frame FUV-to-NIR SEDs present an increase of the rest-frame NIR flux, which is well known to correlate with M$_{*}$ \citep{Kauffmann1998}, and a reddening of the UV/optical part of the spectrum. No significant variations are found in the shape of the FIR emission, but the IR luminosity increases proportionally with M$_{*}$. The top right panel of Figure \ref{sed_mass_beta_model} shows the four SEDs and templates derived from the stacking analysis as a function of $\beta_{\mathrm{UV}}$. Their rest-frame FUV-to-NIR SEDs have higher rest-frame NIR flux for redder $\beta_{\mathrm{UV}}$, and the slope of the UV spectrum follows the mean $\beta_{\mathrm{UV}}$ of the stacked population. The FIR emissions present similar shapes with an increase of $L_{\mathrm{IR}}$ for redder galaxies, except for the bluest $\beta_{\mathrm{UV}}$, which give an IR peak strongly shifted to longer wavelengths. Blue galaxies are very faint on FIR, and we suggest that our FIR stacking analysis on LBG-$\beta$1 is very uncertainty. Besides, if this IR behavior is real we should see it on LBG-$M\beta$1 stacking analysis, and this is not happening. 

The bottom left panel of Figures \ref{sed_mass_beta_model} illustrates the six SEDs and templates derived from the LBG-$M\beta$1 stacking analysis. As a reminder, we split the sample into two large bins of M$_{*}$, and split each of these bins into three bins of $\beta_{\mathrm{UV}}$. Their rest-frame FUV-to-NIR SEDs seem to subdivide into two different sets of templates, high and low M$_{*}$, with variations in the UV part of the SED. No variations in the shape of the FIR part are found, except for an increase of the IR luminosity with M$_{*}$ and $\beta_{\mathrm{UV}}$. The bottom right panel of Figure \ref{sed_mass_beta_model} shows the ten SEDs and templates derived from the LBG-$M\beta$2 stacking analysis. In this case, we split the sample in two large bins of $\beta_{\mathrm{UV}}$, and each of them in five bins of M$_{*}$. Their rest-frame FUV-to-NIR SEDs present similar shapes in comparison with the stacking analysis as a function of M$_{*}$: an increase of the rest-frame NIR flux and a reddening of UV/optical part of the spectrum for massive LBGs. However, the split into $\beta_{\mathrm{UV}}$, red and blue bins, gives access to two templates for each M$_{*}$, with similar NIR shapes, but in a strongly different UV part of the spectrum. The FIR shape does not present a large variety.

\section{Dust attenuation}\label{att}

We analyzed in AM16 how the dust attenuation varies as a function of $\beta_{\mathrm{UV}}$ and $M_{*}$ for our LBG sample by means of the IRX. At that time, we had access to the full FIR/submillimeter part of the spectrum to compute the average L$_{\mathrm{IR}}$. However, the $\beta_{\mathrm{UV}}$ and $M_{*}$ values associated with each stacked LBG SEDs were derived by averaging over the individual values of each LBG inside the stacked bin. The new rest-frame FUV-to-FIR stacking analysis allows us to derive the  $\beta_{\mathrm{UV}}$ and $M_{*}$  directly from the stacked LBG SEDs. We then reviewed the information in the IRX-$\beta_{\mathrm{UV}}$ and IRX-$M_{*}$ planes using the results obtained from the rest-frame FUV-to-FIR SED-fitting analysis on the stacked LBGs SEDs.

\subsection{Dust attenuation as a function of $\beta_{\mathrm{UV}}$}

\begin{figure}[h]
   \centering
   \includegraphics[width=\hsize]{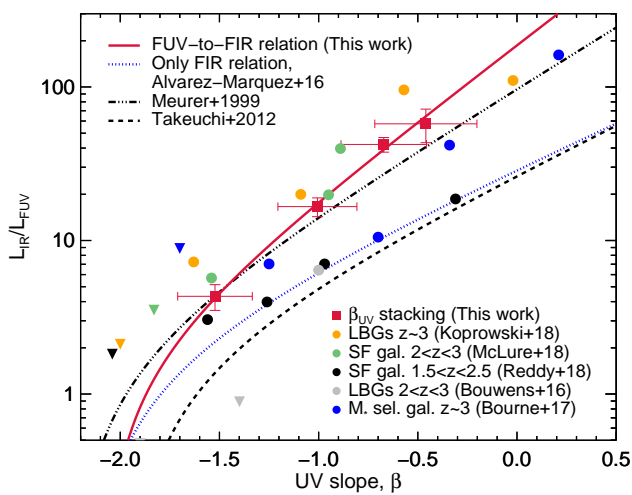}
   \caption{IRX-$\beta_{\mathrm{UV}}$ diagram. Lines show different IRX-$\beta_{\mathrm{UV}}$ relations: the well-known local starburst galaxies calibration (M99; triple-dot-dashed line), the aperture correction of M99 relation by \citeauthor{Takeuchi2012} (2012; dashed line), and the relation derived from a FIR stacking analysis over the same LBG sample (AM16; dotted blue line). In addition, we plot the IRX-$\beta_{\mathrm{UV}}$ results from LBGs at redshifts, $z\sim3$ (\citealt{Koprowski2018}, orange dots) and $2<z<3$ (\citealt{Bouwens2016}, gray dots), star-forming galaxies at redshifts, $2<z<3$ (\citealt{McLure2018}, green dots) and $1.5<z<2.5$ (\citealt{Reddy2018}, black dots), and a mass selected sample at redshift $z\sim3$ (\citealt{Bourne2017}, blue dots). The downward triangles represent upper limits. } 
   \label{IRX_beta}
   \end{figure}

Figure \ref{IRX_beta} shows the IRX-$\beta_{\mathrm{UV}}$ plane, where we plot the results from our rest-frame FUV-to-FIR stacking analysis as a function of $\beta_{\mathrm{UV}}$ (LBG-$\beta$). This is compared with the well-known IRX-$\beta_{\mathrm{UV}}$ relation calibrated from local starburst galaxies (M99) and the relation recomputed by \cite{Takeuchi2012} using the same sample and new photometric data from AKARI and GALEX. We also include our previous IRX-$\beta_{\mathrm{UV}}$ relation derived from a FIR stacking analysis over the same LBG sample (AM16). In addition, we plot the IRX-$\beta_{\mathrm{UV}}$ results from LBGs at redshifts, $z\sim3$ \citep{Koprowski2018} and $2<z<3$ \citep{Bouwens2016}, star-forming galaxies at redshifts, $2<z<3$ \citep{McLure2018} and $1.5<z<2.5$ \citep{Reddy2018}, and a mass selected sample at redshift $z\sim3$ \citep{Bourne2017}. 

The IRX-$\beta_{\mathrm{UV}}$ results from the FUV-to-FIR (this work) and only FIR (AM16) stacking analyses disagree, even if both are computed with the same LBG sample. The FUV-to-FIR results gives  an IRX value that is $\sim$2-3 times higher for a given $\beta_{\mathrm{UV}}$ than the FIR values. The difference resides on the method used to derive the $\beta_{\mathrm{UV}}$ in each of the analyses. In Appendix \ref{beta_comparison}, we demonstrate that $\beta_{\mathrm{UV-power}}$ (AM16) are biased to redder values in comparison with $\beta_{\mathrm{UV-SED}}$ (this work, see Section \ref{slope_uvluminosity}). If the IRX-$\beta_{\mathrm{UV}}$ results from the FIR analysis are corrected by the relation between the $\beta_{\mathrm{UV-power}}$ and $\beta_{\mathrm{UV-SED}}$, obtained in Appendix \ref{beta_comparison}, they are in agreement with the IRX-$\beta_{\mathrm{UV}}$ relation obtained from the FUV-to-FIR analysis. \cite{Koprowski2018} have also confirmed that discrepancies in the IRX-$\beta_{\mathrm{UV}}$ relation are due to bias in the method used to determine the $\beta_{\mathrm{UV}}$. These authors suggested that these inconsistencies are driven by scatter in measured values of $\beta_{\mathrm{UV}}$ from limited photometry that serves to artificially flatten the  IRX-$\beta_{\mathrm{UV}}$ relation. They obtained that the scatter is significantly reduced by determining $\beta_{\mathrm{UV}}$ from SED-fitting analysis, which is the same conclusion presented in Appendix \ref{beta_comparison}. 

Our FUV-to-FIR stacking and SED-fitting analysis results as a function of $\beta_{\mathrm{UV}}$ follow the IRX-$\beta_{\mathrm{UV}}$ local relation (M99) within the uncertainties. These results are in agreement with previous stacking analyses done on star-forming galaxies at high-$z$. \cite{Magdis2010c} found that the dust-corrected UV-SFR derived from the M99 relation presents a good match with the FIR and radio SFR estimators for stacked IRAC spectropically confirmed LBGs at $z\sim3$. \cite{Koprowski2018}, by stacking a sample of $\sim$4000 LBGs at redshifts, $3<z<5$, have concluded that LBGs are consistent with the M99 IRX-$\beta_{\mathrm{UV}}$ relation and do not present a redshift evolution. Similar results were obtained by \cite{McLure2018}, who studied a sample of star-forming galaxies at redshifts, $2<z<3$, from a deep ALMA 1.3 mm continuum data. Additional stacking analysis of spectroscopically confirmed $z\sim2$ LBGs in FIR \citep{Reddy2012}, $z\sim4$ LBGs in the radio continuum (1.4GHz; \citealt{To2014}), and direct detections of LBGs in PACS at lower redshift ($z\sim1$; \citealt{Oteo2013a, Burgarella2011}) also agree with the M99 relation. \cite{Bourne2017} have used a deconfusion methodology to reach below the confusion limit maps (SCUBA2 and Herschel) for massive galaxies at redshifts, $0.5<z<6$, detected in the CANDELS fields (AEGIS, COSMOS, and UDS). These authors have obtained similar results as for local starburst galaxies in M99, although their results deviate to higher IRX values at high stellar masses, which is similar to the analysis presented in Sect. \ref{dispersion}. However, \cite{Reddy2018} have used $\sim3500$ star-forming galaxies at redshifts, $1.5<z<2.5$, and have concluded that the M99 IRX-$\beta_{\mathrm{UV}}$ local relation overpredicts the IRX by a factor of $\sim3$ at a given $\beta_{\mathrm{UV}}$. \cite{Fujimoto2016} and \cite{Bouwens2016} used ALMA observations of LBGs at $z\sim2$-3, and also showed that the M99 relation overpredicts the IRX. The ALMA observations contain small samples that could make it difficult to constrain the IRX-$\beta_{\mathrm{UV}}$ relation statistically. In particular, \cite{Bouwens2016} stacked a LBG sample at $z=2-3$ in bins of $M_{*}$ and $\beta_{\mathrm{UV}}$. Their results suggested an IRX-$\beta_{\mathrm{UV}}$ values below the M99 relation, but the high $M_{*}$ bin (log(M$_{*}$ [M$_{\odot}$]) $>$ 9.75) is in agreement within the uncertainties.

\subsection{Dust attenuation as a function of $M_{*}$}

\begin{figure}[h]
   \centering
   \includegraphics[width=\hsize]{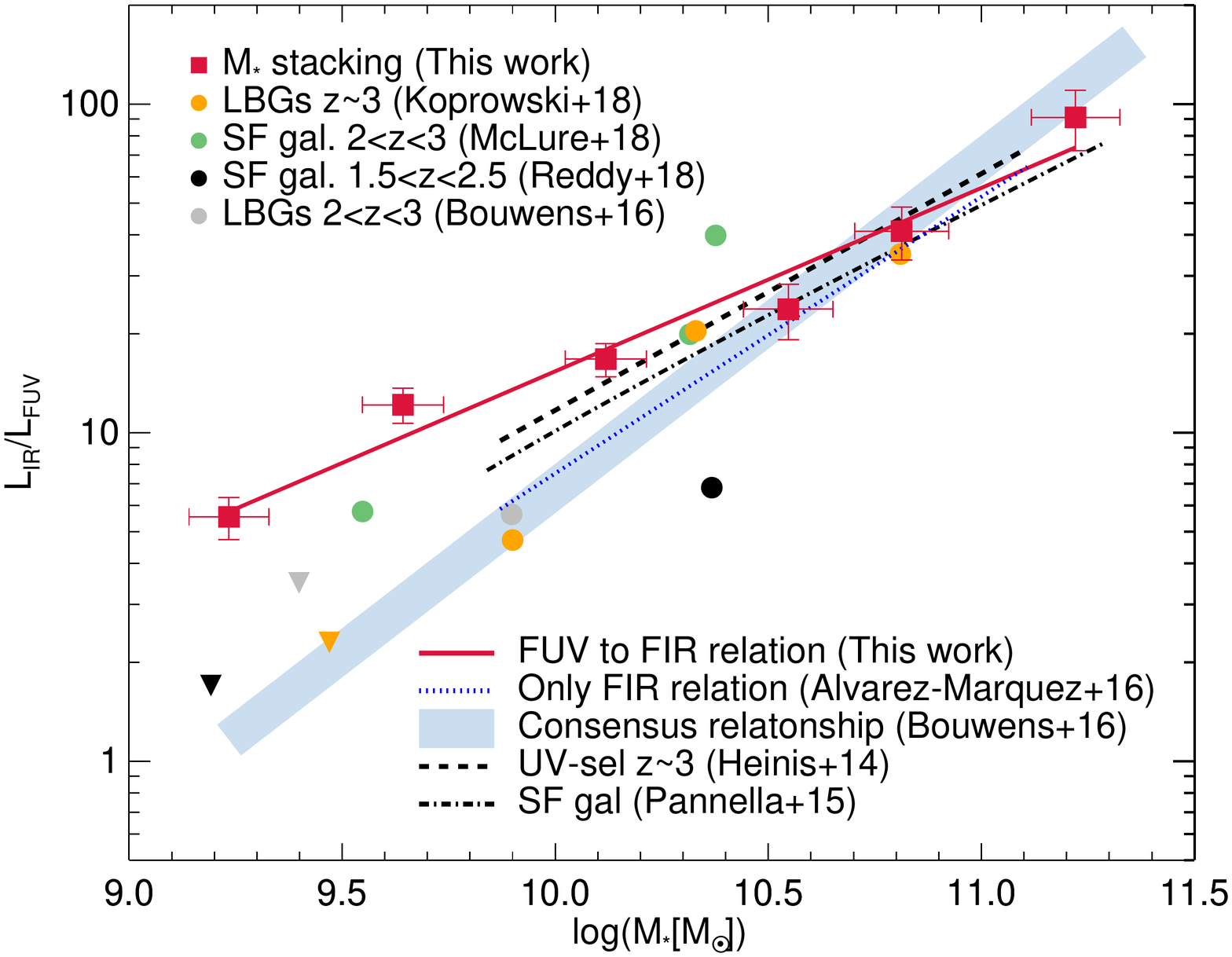}
   \caption{IRX-$M_{*}$ diagram. Our rest-frame FUV-to-FIR stacking analysis as a function of $M_{*}$ (LBG-$M$) results and the power-law fit are shown by red squares and solid line. The blue line presents the consensus relationship for star-forming galaxies at redshifts $2<z<3$ \citep{Bouwens2016}, and defined by a combination of different works done in the literature \citep{Reddy2010, Whitaker2014, Alvarez-marquez2016}. This relation includes that previously derived from a FIR stacking analysis over our LBG sample (AM16, dotted blue line). We overplot the relations obtained from a stacking analysis of UV-selected galaxies at redshift $z\sim3$ (\citealt{Heinis2014}, dashed line) and a complete sample of star-forming galaxies up to $z\sim4$ (\citealt{Pannella2015}, dot-dashed line). We also show the IRX-$M_{*}$ results from LBGs at redshifts, $z\sim3$ (\citealt{Koprowski2018}, orange dot) and $2<z<3$ (\citealt{Bouwens2016}, gray dot), and star-forming galaxies at redshifts $2<z<3$ (\citealt{McLure2018}, green dots) and $1.5<z<2.5$ (\citealt{Reddy2018}, black dots).  The downward triangles represent upper limits.} 
   \label{IRX_mass}
   \end{figure}

Figure \ref{IRX_mass} shows the IRX-$M_{*}$ plane, where the results from our rest-frame FUV-to-FIR stacking analysis as a function of $M_{*}$ (LBG-$M_{*}$) are plotted. These are compared with the relationship presented by \cite{Bouwens2016} for star-forming galaxies at redshifts $2<z<3$ and defined from the combination of different works done in the literature \citep{Reddy2010, Whitaker2014, Alvarez-marquez2016}. This relation includes that previously derived from a FIR stacking analysis over our LBG sample (AM16). We overplot the relations obtained from a stacking analysis of UV-selected galaxies at redshift $z\sim3$ \citep{Heinis2014} and a complete sample of star-forming galaxies up to $z\sim4$ \citep{Pannella2015}. We also show the IRX-$M_{*}$ results from LBGs at redshifts, $z\sim3$ \citep{Koprowski2018} and $2<z<3$ \citep{Bouwens2016}, and star-forming galaxies at redshifts, $2<z<3$ \citep{McLure2018} and $1.5<z<2.5$ \citep{Reddy2018}. The IRX results from the FUV-to-FIR stacking analysis show a good agreement at the highest stellar mass bins (log(M$_{*}$ [M$_{\odot}$]) > 10). However, the IRX results disagree at low stellar masses bins (log(M$_{*}$ [M$_{\odot}$]) < 10) presenting larger IRX values for a given M$_{*}$.

We suggest that an incompleteness of our LBG sample in terms of M$_{*}$ might be the origin of the IRX-M$_{*}$ discrepancies. If M$_{*}$ is proportional to the rest-frame NIR emission \citep{Kauffmann1998}, the incompleteness of our LBG sample at $z\sim3$ in terms of M$_{*}$ is proportional to the detectability of our LBGs at IRAC 3.6 $\mu$m observations. The mean fluxes of the LBG population with stellar masses, log(M$_{*}$ [M$_{\odot}$]) < 10, are between 1.2 to 3.6 $\mu$Jy (see Appendix \ref{table_stak_ir}). \cite{Ilbert2010} showed that the i-band photometric redshift catalog (\citealt{Ilbert2009}, version 2.0), used to build our LBG sample in AM16 is 90\% complete at 5$\mu$Jy and 50\% complete at 1$\mu$Jy in the IRAC 3.6 $\mu$m observations. Then, the low stellar mass bins (log(M$_{*}$ [M$_{\odot}$]) < 10) of LBG-$M$ stacking analysis are between 80\% to 50\% complete in term of M$_{*}$. 

However, there are some indications in the literature that the origin of the departure might not be due to M$_{*}$ bias. For instance, \cite{Whitaker2017} showed that the region corresponding to galaxies at log(M$_{*}$ [M$_{\odot}$]) $<$ 10 and high IRX values is populated by galaxies. This effect does not seem to be at z = 0, however. It could also be noticed in \cite{Heinis2014} that there is a trend for galaxies with low-$L_{\mathrm{UV}}$ to have larger IRX. We can wonder whether this effect does not hide the fact that these low-$L_{\mathrm{UV}}$ only trace dust-free stars in galaxies that might be otherwise dusty (meaning that most of the stars do not contribute to the UV luminosity). Finally, \cite{Spitler2014} studied a sample of 57 galaxies with log(M$_{*}$ [M$_{\odot}$]) $<$ 10.6, and they find that massive star-forming galaxies can be found at all $M_{\mathrm{UV}}$. So, we propose that the trend we observe at log(M$_{*}$ [M$_{\odot}$]) $<$ 10 in Figure \ref{IRX_mass} might be real and not due to an observational bias. The impact of this hypothesis will be further studied in Bogdanoska et al. (in prep.). In Section \ref{dispersion}, when LBGs are stacked as a function of M$_{*}$ and $\beta_{\mathrm{UV}}$, a population of LBGs appear to locate at the same region of the IRX-M$_{*}$ diagram than \cite{Whitaker2017}. If this population is removed to perform the stacking analysis and derive the mean IRX-M$_{*}$ relation (light red filled square in Figure \ref{comp_IRX_beta_mass_dispersion_chap5}), the results are closer to the IRX-M$_{*}$ relations from \cite{Pannella2015} and \cite{Heinis2014}, and the star-forming galaxies at $2<z<3$ stacked in ALMA observation \citep{McLure2018}.

   \begin{figure}[h]
   \centering
   \includegraphics[width=\hsize]{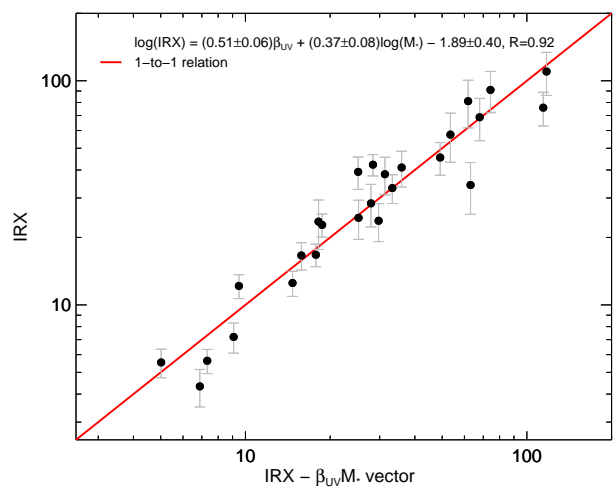}
   \caption{Empirical IRX calibration by combination the $\beta_{\mathrm{UV}}$ and M$_{*}$. The plane fit in the 3D IRX-$\beta_{\mathrm{UV}}$-M$_{*}$ diagram provides us with the equation $\log(IRX) = (0.51\pm0.06)\beta_{\mathrm{UV}} + (0.37\pm0.08)\log(M_{*} [M_{\odot}]) - 1.89\pm0.40$, which can be used to correct the dust attenuation when $\beta_{\mathrm{UV}}$ and M$_{*}$ are known. The y-axis corresponds to the IRX obtained from SED-fitting analysis over the stacked LBGs SEDs. The x-axis corresponds to the IRX calculated from the plane fit equation. The black points indicate the results from the SED-fitting and stacking analysis as a function of $\beta_{\mathrm{UV}}$ and $M_{*}$, and the combination of the two (LBG-$\beta$, LBG-$M$, LBG-$M\beta$1, and LBG-$M\beta$2). The red line indicates the 1-to-1 relation.}
   \label{IRX_betamass_2dplanefit}
   \end{figure}
   
\subsection{Dispersion on the IRX-$M_{*}$ and IRX-$\beta_{\mathrm{UV}}$ planes}\label{dispersion}

 \begin{figure*}[h]
\begin{minipage}{.4999\textwidth}
\includegraphics[width=\textwidth]{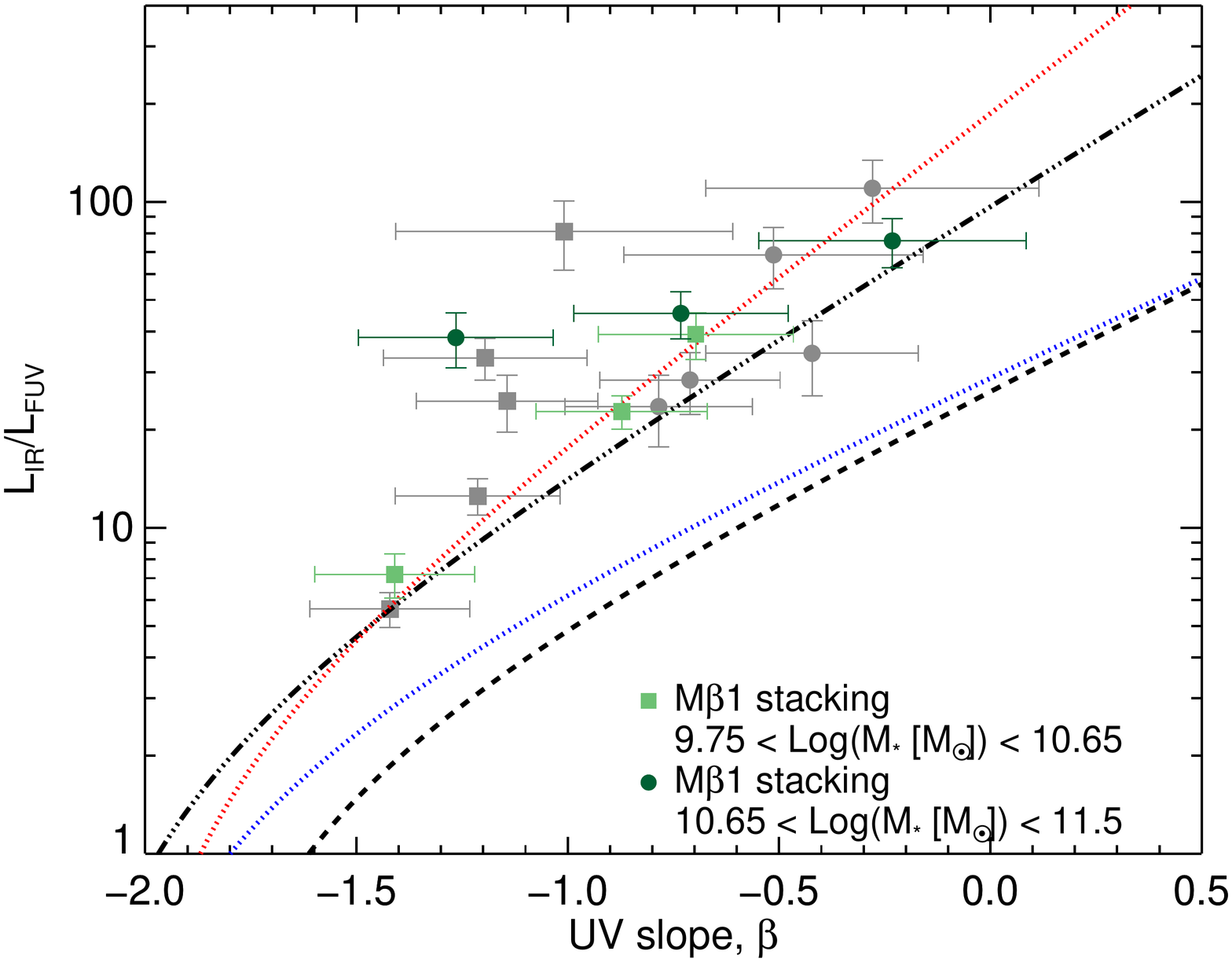}
\end{minipage}
\begin{minipage}{.4999\textwidth}
\includegraphics[width=\textwidth]{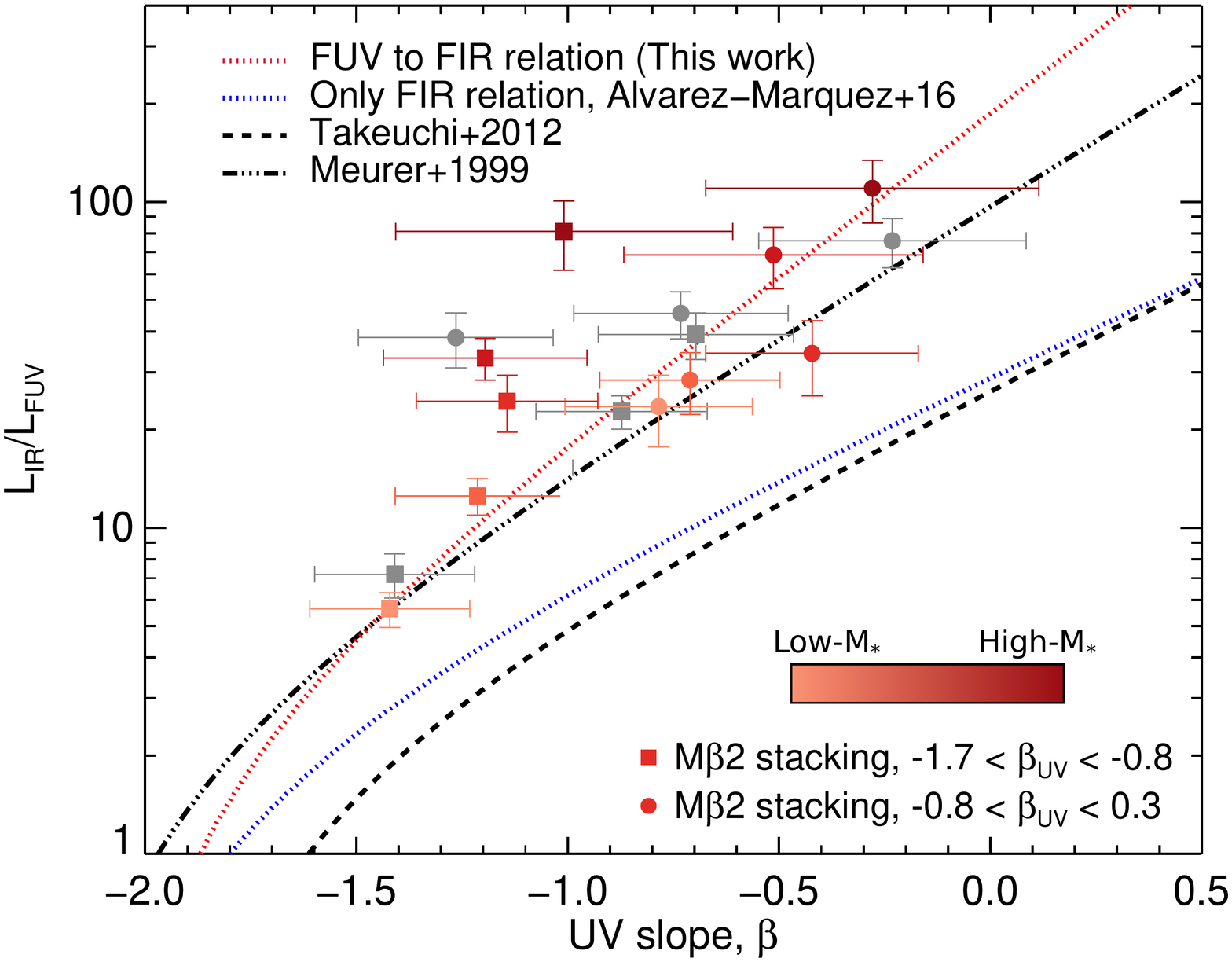}
\end{minipage}
\begin{minipage}{.4999\textwidth}
\includegraphics[width=\textwidth]{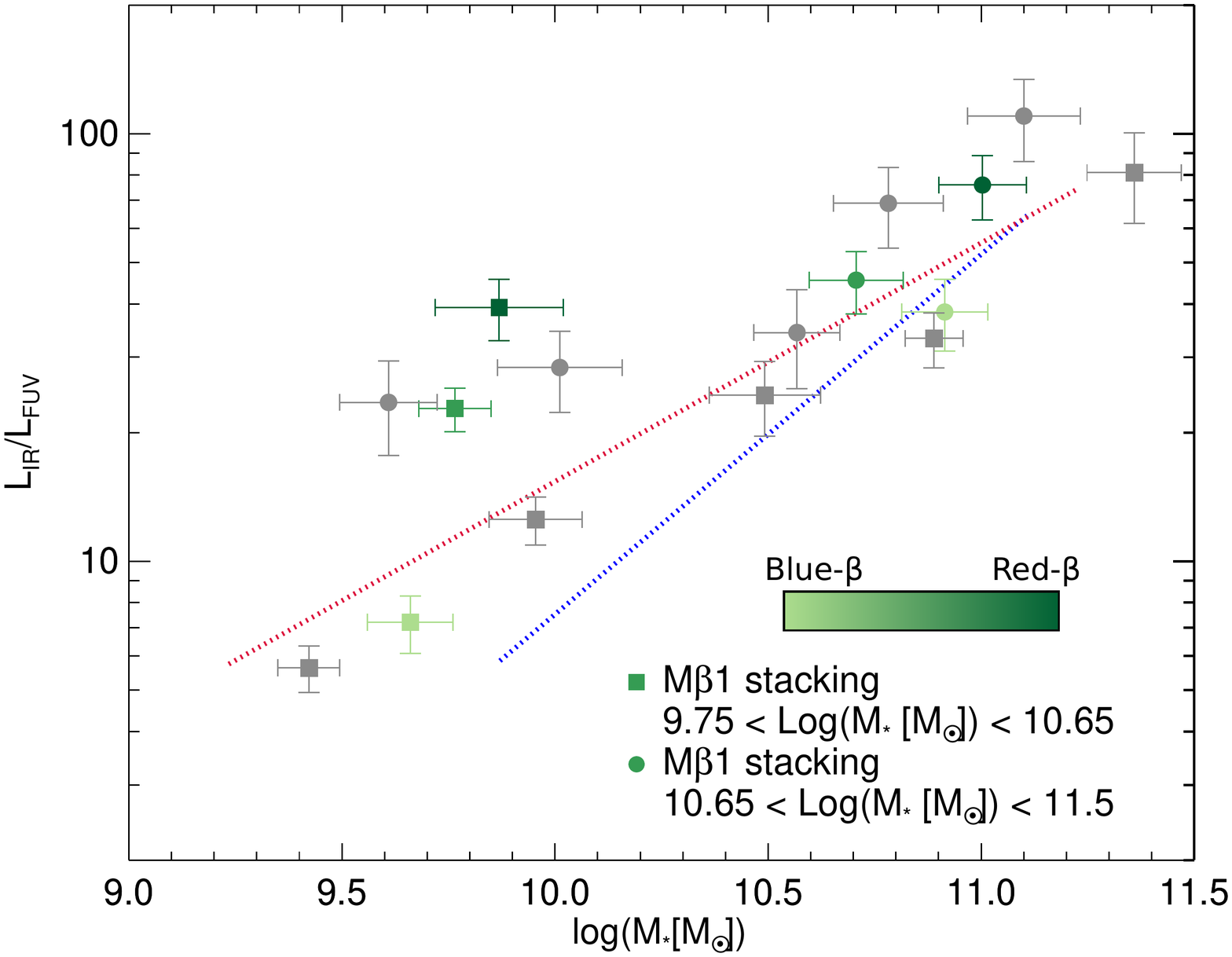}
\end{minipage}
\begin{minipage}{.4999\textwidth}
\includegraphics[width=\textwidth]{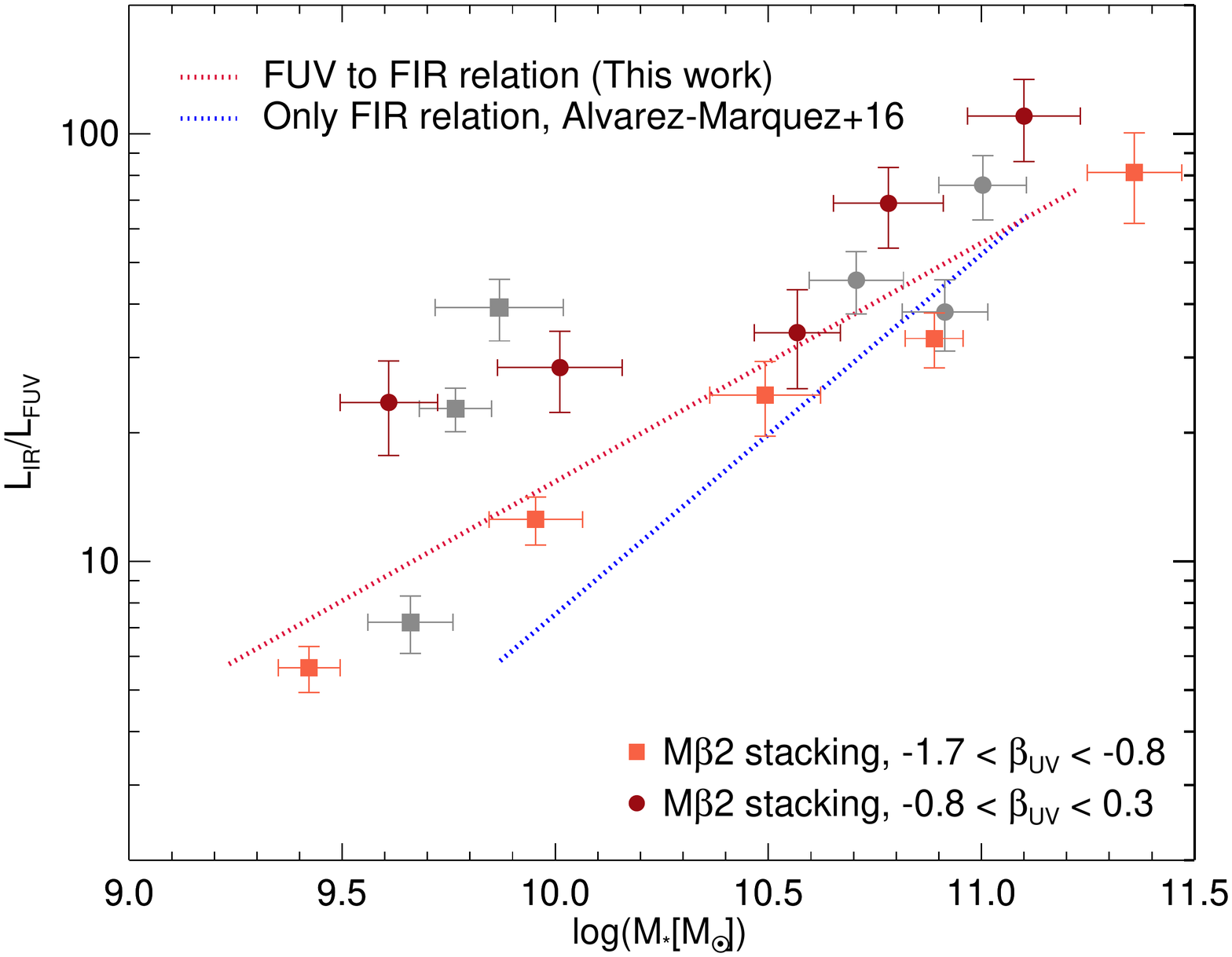}
\end{minipage}
      \caption{LBG-$M\beta$1 and  LBG-$M\beta$2 results in the IRX-$\beta_{\mathrm{UV}}$ (top panels) and IRX-$M_{*}$ (bottom panels) planes. Top left panel: the LBG-$M\beta$1 are shown in green; the tonalities represent the two different bins in $M_{*}$, $9.75 < \log(M_{*}[M_{\odot}]) < 10.65$ (light green filled square) and $10.65 < \log(M_{*}[M_{\odot}]) < 11.5$ (dark green filled circles). Top right panel: the LBG-$M\beta$2 results are shown in red; the tonalities represent the increase of the M$_{*}$ from 9.75 (light red) to 11.50 (dark red). The filled squares represent the bluer $\beta_{\mathrm{UV}}$ bin and the filled circles the redder $\beta_{\mathrm{UV}}$ bin. Bottom left panel: LBG-$M\beta$1 are shown in green; the tonalities represent the increase of the $\beta_{\mathrm{UV}}$ from -1.7 (light green) to 0.5 (dark green). The filled squares represent the low M$_{*}$ bin and the filled circles the high M$_{*}$ bin. Bottom right panel: The LBG-$M\beta$2 results are shown in red; the tonalities represent the two different bins in $\beta_{\mathrm{UV}}$,  $-1.7 < \beta_{\mathrm{UV}} < -0.5$ (light red filled square) and $-0.5 < \beta_{\mathrm{UV}} < 0.5$ (dark red filled circles). The gray points in all diagrams represent the LBG-$M\beta$ results that are not highlighted in color.}
         \label{comp_IRX_beta_mass_dispersion_chap5}
   \end{figure*}

We use our stacking analyses (LBG-$M\beta$) to investigate the dispersion on the IRX-$\beta_{\mathrm{UV}}$ and IRX-$M_{*}$ planes, and the validity of the IRX-$\beta_{\mathrm{UV}}$ and IRX-$M_{*}$ mean relations. On the LBG-$M\beta$ stacking analyses, the LBG sample is split on the ($M_{*}$, $\beta_{\mathrm{UV}}$) plane in two different ways. For LBG-$M\beta$1, the LBG sample is divided into two bins of M$_{*}$, and each of these is split as a function of $\beta_{\mathrm{UV}}$. And for LBG-$M\beta$2, it is divided into two bins of $\beta_{\mathrm{UV}}$ and each of these is split as a function of M$_{*}$. Figure \ref{comp_IRX_beta_mass_dispersion_chap5} shows the results from the LBG-$M\beta$ stacking analyses in the IRX-$\beta_{\mathrm{UV}}$ and IRX-$M_{*}$ planes. The results suggest large dispersion on both planes, as a consequence of the M$_{*}$ evolution in the IRX-$\beta_{\mathrm{UV}}$ plane and the $\beta_{\mathrm{UV}}$ in the IRX-$M_{*}$ plane. On the one hand, high stellar mass LBGs present higher IRX values than the mean IRX-$\beta_{\mathrm{UV}}$ relation. Similar results were reported by \cite{Bourne2017} in the IRX-$\beta_{\mathrm{UV}}$ plane. These authors showed an evolution of the IRX-$\beta_{\mathrm{UV}}$ relation from low to high stellar masses using a sample of massive galaxies at redshifts $0.5<z<6$. Some studies of IR-selected galaxy samples have found IRX-$\beta_{\mathrm{UV}}$ values located above M99 relation \citep{Casey2015, Oteo2013, Buat2015}, which suggest that the criterion to select the sample has a strong impact on the mean dust attenuation of a galaxy population. On the other hand, redder LBGs tend to have higher IRX values than the mean IRX-$M_{*}$ relation. In particular, a population of LBGs with $\beta_{\mathrm{UV}} < -1$ and stellar masses (log(M$_{*}$ [M$_{\odot}$]) $<$ 10) shows large IRX values that flatten the mean IRX-M$_{*}$ presented in Figure \ref{IRX_mass}.

The dependence of M$_{*}$ in the IRX-$\beta_{\mathrm{UV}}$ plane and $\beta_{\mathrm{UV}}$ in the IRX-M$_{*}$ plane makes us think about the utility and efficiency of providing a new empirical IRX relation combining the $\beta_{\mathrm{UV}}$ and M$_{*}$. It may reduce the uncertainty to derive the dust attenuation for galaxies for which only UV/optical/NIR observations are available. We use the results from the SED-fitting and stacking analysis as a function of  $\beta_{\mathrm{UV}}$ and $M_{*}$, and the combination of the two (LBG-$\beta$, LBG-$M$, LBG-$M\beta$1, and LBG-$M\beta$2) to attempt performing a plane fit in the 3D IRX-$\beta_{\mathrm{UV}}$-M$_{*}$ diagram. We obtain a mean relation equal to $\log(IRX) = (0.51\pm0.06)\beta_{\mathrm{UV}} + (0.37\pm0.08)\log(M_{*} [M_{\odot}]) - 1.89\pm0.40$. Figure \ref{IRX_betamass_2dplanefit} shows the dispersion between the IRX of each stacked LBGs obtained from SED-fitting analysis, and the IRX derived from the mean IRX-$\beta_{\mathrm{UV}}$-M$_{*}$ relation.

   \begin{figure}[h]
   \centering
   \includegraphics[width=\hsize]{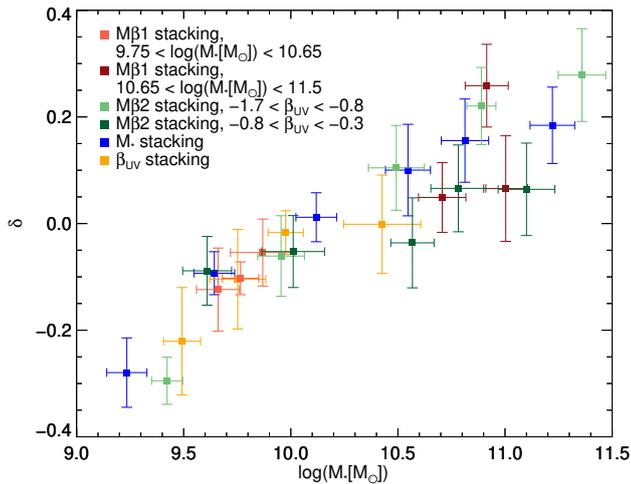}
      \caption{Change of the slope of the dust attenuation law with respect to Calzetti ($\delta$) as a function of stellar mass. The different colors show each stacking analysis as a function of $\beta_{\mathrm{UV}}$ and $M_{*}$, and the combination of both parameters (LBG-$\beta$, LBG-$M$, LBG-$M\beta$1, and LBG-$M\beta$2). This suggests a correlation between the stellar mass of a galaxy and the shape of the dust attenuation curve, which is steeper for LBGs with $\log(M_{*} [M_{\odot}]) < 10.25$ and grayer for LBGs with $\log(M_{*} [M_{\odot}]) > 10.25$ than Calzetti's law.} 
         \label{delta_stellarmass_beta}
   \end{figure}

\subsection{Shape of the dust attenuation curve for LBGs at $z\sim3$}\label{delta_att}

The Calzetti law is frequently assumed when studying star-forming galaxies at high-$z$ (e.g., \citealt{Bouwens2009, Schaerer2013, Finkelstein2015}). From observations in the local universe, it is well known that dust attenuation curves are not universal across galaxies. Most sight lines in the Milky Way (MW) have a strong extinction bump near 2175 $\AA$ \citep{Stecher1965, Cardelli1989}. However, the feature is weaker in the Large Magellanic Cloud (LMC; \citealt{Koornneef1981}) and absent in the Small Magellanic Cloud (SMC; \citealt{Prevot1984}). This diversity is also reflected by systematic changes in the FUV slope, which gets steeper from MW to SMC. \cite{Calzetti1994} deduced a mean dust attenuation curve from a sample of 39 local UV-bright starburst galaxies. Their measurements are characterized by a grayer FUV slope than both MW and LMC extinction curves and an absence of the 2175 $\AA$ absorption feature. At higher redshifts, a variety can be also seen. \cite{Buat2011, Buat2012} found for a UV-selected galaxy at $z>1$ a clear present of 2175 $\AA$ bump and evidence for a steeper rise of the dust attenuation curve in comparison to the Calzetti law. Other studies support this model, demonstrating that the 2175 $\AA$ bump is typically visible in normal star-forming galaxies (e.g., \citealt{Burgarella2005, Noll2007,Conroy2010}). Other authors have also suggested that dust attenuation curves vary significantly at any redshift (e.g., \citealt{KriekConroy2013, Zeimann2015, Salmon2016, LoFaro2017}). This diversity in the shape of the dust attenuation curves could be produced by the differences in the dust grain properties, the relative geometrical distribution of stars and dust within a galaxy, the line-of-sight galaxy orientation, and the galaxy type \citep{WittGordon2000, Pierini2004, Tuffs2004, Chevallard2013}.

The results from the SED-fitting analysis over our stacked LBGs SEDs suggest a large variation in the shape of the dust attenuation curve across the LBG population at $z\sim3$. We obtain a range of $\delta$'s from -0.3 to 0.3 (see Appendix \ref{physical_param}), which produce dust attenuation curves steeper and grayer than Calzetti law ($\delta=0$), respectively. Similar results were reported by \cite{Salmon2016}, who found a range of $\delta$'s from -0.5 to 0.2 in star-forming galaxies at $z\sim1.5-3$. Figure \ref{delta_stellarmass_beta} illustrates the dependence of $\delta$ as a function of M$_{*}$ for the stacking analysis as a function of $\beta_{\mathrm{UV}}$ and $M_{*}$, and the combination of these both parameters (LBG-$\beta$, LBG-$M$, LBG-$M\beta$1, and LBG-$M\beta$2). We obtain that $\delta$ is well correlated with the M$_{*}$. The low stellar mass population of LBGs ($\log(M_{*} [M_{\odot}]) < 10.25$) favors steeper dust attenuation curves than the Calzetti's law, and the large stellar mass population of LBGs ($\log(M_{*} [M_{\odot}]) > 10.25$) favors grayer Calzetti's law. A similar trend with the stellar mass is seen in the local Universe by \cite{Salim2018}; these authors have suggested that the apparent dependence with the stellar mass is due to a relation between the stellar mass and optical opacity. They conclude that opaque galaxies have shallower dust attenuation laws. \cite{Cullen2018} studied a sample of star-forming galaxies at redshifts, $3<z<4$, and stellar mass range, $8.2 < \log(M_{*} [M_{\odot}]) < 10.6$. Their mean FUV-to-optical dust attenuation curve is in agreement with Calzetti's, and gets steeper at lower masses. However, \cite{Zeimann2015} performed an analysis of star-forming galaxies selected by their rest-frame optical emission lines at redshifts, $1.9 < z < 2.35$, and mass range, $7.2 < \log(M_{*} [M_{\odot}]) < 10.2$. They found grayer dust attenuation curves than Calzetti's, which differs from our findings.

 \begin{figure}[h]
   \centering
   \includegraphics[width=\hsize]{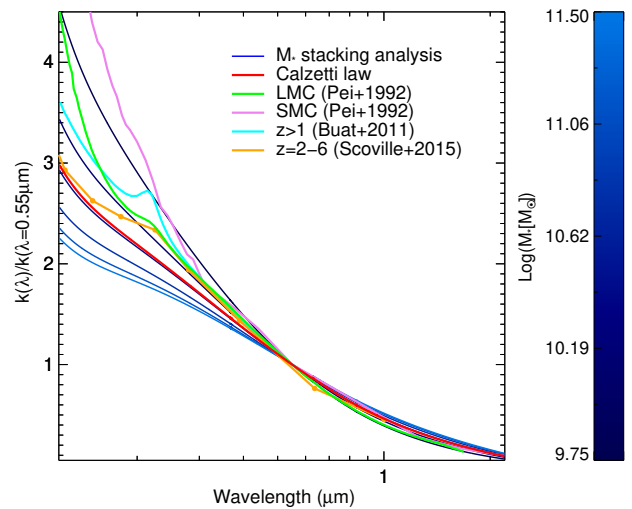}
      \caption{Dust attenuation curves derived from the SED-fitting analysis over the stacked LBGs SEDs obtained from the stacking analysis as a function of M$_{*}$. These curves are color coded as a function of  M$_{*}$: light blue represents larger stellar masses which are grayer than Calzetti's law (red line), and  dark blue corresponds to lower stellar masses which are steeper than Calzetti's law. As a comparison, we also show the LMC and SMC dust extinction curve \citep{Pei1992} and two additional attenuation curves derived from UV-selected galaxies at $z>1$ \citep{Buat2011} and star-forming galaxies at $z\sim2$ to 6 \citep{Scoville2015}.} 
         \label{dust_att_curve_stellarmass}
  \end{figure} 

Figure \ref{dust_att_curve_stellarmass} illustrates the derived dust attenuation curves in the stacked LBGs SEDs as a function of M$_{*}$ (LBG-M). This figure shows how dust attenuation curves get steeper when the stellar mass decreases, up to the lowest stellar mass bin that presents a slope closer to the SMC dust attenuation curve \citep{Pei1992}.  Other mean dust attenuation laws derived from UV-selected galaxies at $z > 1$ \citep{Buat2011} and star-forming galaxies at $z\sim2$ to 6 in the COSMOS field \citep{Scoville2015} are also compared; they are in agreement with the derived dust attenuation curves for our stacked LBGs SEDs.
  
We note in our stacked LBGs SEDs that some of the z-band ($\lambda_{rest} \sim 2200\AA$) data points have lower observed flux than its consecutive data points (i-band and Y-band). This might be a consequence of a larger dust attenuation owing to the coincidence with the UV bump at 2175$\AA$. As we commented in Section \ref{initial_param_sedfitting}, the spectral resolution of the FUV-to-FIR stacked LBGs SEDs is not sufficient to constrain the amplitude and shape of the UV bump, and its amplitude has been set to zero in our SED-fitting analysis. However, the presence or not of a UV bump in the dust attenuation law might modify the derived $\delta$ values. We performe an additional SED-fitting analysis with the amplitude of the UV bump as a free parameter to check the validity of the derived $\delta$ values. We obtain a mean differences, $\langle \delta_{D_{\lambda_{o}, \gamma, E_{bump}-free}} - \delta_{D_{\lambda_{o}, \gamma, E_{bump}=0}} \rangle = -0.013 \pm 0.008$, which are within the derived $\delta$'s uncertainties of the initial SED-fitting analysis.

\subsection{IRX-$\beta_{\mathrm{UV}}$ plane and the slope of the dust attenuation curve}

We previously reported a large dispersion in the IRX-$\beta_{\mathrm{UV}}$ plane associated with a M$_{*}$ variations. We also obtain a diversity of dust attenuation curves along our LBG sample, which correlates with the M$_{*}$. We check in this section if the dispersion on the IRX-$\beta_{\mathrm{UV}}$ plane is a consequence of a variety in M$_{*}$ or in the shape of the dust attenuation curve. A few recent studies showed that the diversity in the shape of the dust attenuation curves has a strong impact on shaping the IRX-$\beta_{\mathrm{UV}}$ plane (e.g., \citealt{Salmon2016}; \citealt{LoFaro2017}). 

   \begin{figure}[h]
   \centering
   \includegraphics[width=\hsize]{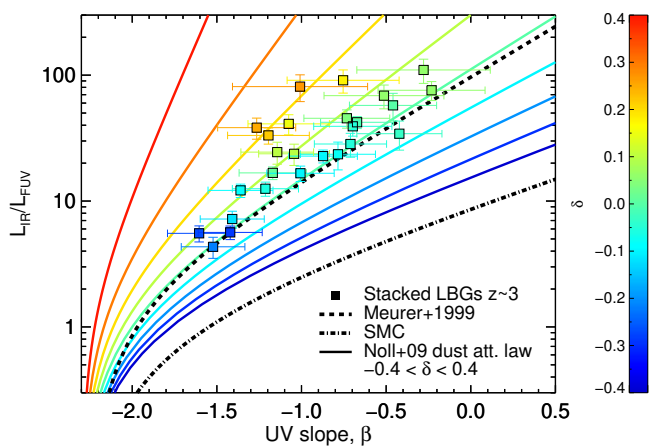}
     \caption{IRX-$\beta_{\mathrm{UV}}$ diagram seeing from a $\delta$ point of view. Continuous lines show the simulation performed by CIGALE with a fixed SFH and metallicity for different dust attenuation curves (see text). The square points represent the results from the SED-fitting analysis and stacking analysis as a function of $\beta_{\mathrm{UV}}$ and $M_{*}$, and the combination of the two (LBG-$\beta$, LBG-$M$, LBG-$M\beta$1, and LBG-$M\beta$2). Both are color coded as a function $\delta$. The dashed and dot-dashed lines represent the classical M99 and SMC \citep{Bouwens2016} relations, respectively} 
         \label{IRX_beta_deltavariation}
  \end{figure}
  
The observed UV part of the spectrum, where the $\beta_{\mathrm{UV}}$ is calculated, depends on different physical properties of a galaxy (e.g., SFH, metallicity,  and dust attenuation law).  If we focus on the dust attenuation law and fix the others properties, when the UV shape of the dust attenuation curve changes the observed UV part of the spectrum would  consequently be modified along with its associated $\beta_{\mathrm{UV}}$. To test this, we use CIGALE in its \textit{SED-modeling} mode to compute synthetic SEDs with a fixed configuration (delayed-SFH with $\tau$=100Myr and age = 300Myr, metallicity = 0.02Z${\odot}$, and fixed IR model), and different dust attenuation laws (-0.4 < $\delta$ < 0.4) and reddening (0 < E(B-V) < 1.5).

Figure \ref{IRX_beta_deltavariation} illustrates the IRX-$\beta_{\mathrm{UV}}$ plane color coded as a function of $\delta$. The continuous lines represent the simulations performed with CIGALE for the different $\delta$ values. Dust attenuation curves steeper than the Calzetti’s law, with $\delta < 0$, are \textbf{located} below the M99 relation, and dust attenuation curves grayer than Calzetti’s law, with $\delta > 0$, are located above. We note that the IRX-$\beta_{\mathrm{UV}}$ relations simulated for each $\delta$ are not unique as we show in this work. This is a consequence of fixing the metallicites and SFH. These relations should be broader if we consider different SFHs and metallicities, as in \cite{Salmon2016} and \cite{LoFaro2017}. We overplot the results from the SED-fitting and stacking analysis as functions of $\beta_{\mathrm{UV}}$ and $M_{*}$ and the combination of both parameters (LBG-$\beta$, LBG-$M$, LBG-$M\beta$1, and LBG-$M\beta$2). Taking into account the uncertainties in the SFH and metallicity, our stacked LBGs are in agreement with our simulations. Our stacked LBGs with associated $\delta\sim0$ follow M99, and when $\delta$ increases the objects present bluer $\beta_{\mathrm{UV}}$ and higher IRX values than M99 relation. We conclude that dust attenuation curves are one of the main drivers that shapes the IRX-$\beta_{\mathrm{UV}}$ plane, following previous studies in other galaxy population (e.g., \citealt{Salmon2016}; \citealt{LoFaro2017}; \citealt{Salim2019}).  

\section{Summary and conclusions}\label{conclusions}

We investigate the full rest-frame FUV-to-FIR emission of LBGs at $z\sim3$ by stacking analysis at the optical ($BVriz$ bands), NIR ($YJHKs$ bands), IRAC (3.6, 4.5, 5.6 and 8.0 $\mu$m), MIPS (24$\mu$m), PACS (100 and 160~$\mu$m), SPIRE (250, 350, and 500~$\mu$m), and AzTEC (1.1mm) observations. We use a subsample of $\sim$17000 LBGs from those previously  selected and characterized in AM16. We split our LBG sample as a function of the single parameters L$_{\mathrm{FUV}}$, $\beta_{\mathrm{UV}}$, and M$_{*}$, and the combination of both $\beta_{\mathrm{UV}}$ and M$_{*}$ in the ($\beta_{\mathrm{UV}}$, M$_{*}$) plane. This allows us to build 30 rest-frame FUV-to-FIR LBGs SEDs at $z\sim3$ and investigate the evolution of their physical properties as a function of the binning parameters.

We use CIGALE, a physically oriented spectral synthesis and SED-fitting code, to analyze our rest-frame FUV-to-FIR stacked LBGs SEDs. The CIGALE code provides us with  the synthetic model spectra that better fit our stacked LBGs SEDs by $\chi^{2}$ minimization and the mean physical parameters that characterize each of the stacked LBG SEDs by applying a Bayesian analysis. After performing exhaustive checks on the validity of the derived physical parameters, we conclude that the SED-fitting analysis can derive fully consistent physical parameters, i.e., $M_{*}$, L$_{\mathrm{IR}}$, A$_{\mathrm{FUV}}$, SFR, and change of the slope of the dust attenuation law with respect to Calzetti  - $\delta$.

We use the stacked LBGs SEDs and their associated modeled spectra to build a library of 30 SEDs and templates of LBGs at $z\sim3$. Thanks to the binning configuration used to perform the stacking analysis, we derive a set of templates with a large variety of physical properties. The library contains templates within an interval of stellar mass, $9.2 < \log(M_{*}\,[M_{\odot}]) < 11.4$, SFR, $20 < SFR\, [M_{\odot}yr^{-1}] < 300$,  $\beta_{\mathrm{UV}}$,  $ -1.8 < \beta_{\mathrm{UV}} < -0.2$, FUV dust attenuation, $1.5 < A_{\mathrm{FUV}}\, [mag] < 4.8$, IR luminosity, $11.2 <  \log(L_{\mathrm{IR}} \,[L_{\odot}]) < 12.7$, and FUV luminosity, $10.4 <  \log(L_{\mathrm{FUV}} \,[L_{\odot}]) < 11.2$. This diversity makes our library very versatile in the sense that our templates fit the physical properties of a large population of star-forming galaxies at $z\sim3$. 

We use the so-called IRX and the mean physical parameters derived for each of the rest-frame FUV-to-FIR stacked SEDs to investigate the dust attenuation properties of LBGs at $z\sim3$. We conclude as follow:

\begin{itemize} 
\item Our LBG sample follows the well-known IRX-$\beta_{\mathrm{UV}}$ calibration from local starburst galaxies (M99), when LBGs are stacked as a function of $\beta_{\mathrm{UV}}$ (LBG-$\beta$), which is in agreement with recent works done on star-forming galaxies at high-z \citep{Koprowski2018,McLure2018,Bourne2017}.

\item Our LBG sample, stacked as a function of M$_{*}$ (LBG-$M$), is in agreement with most of the IRX-$M_{*}$ relations presented in the literature at high stellar masses (log(M$_{*}$ [M$_{\odot}$]) > 10, \citealt{Pannella2015, Heinis2014, Bouwens2016, Alvarez-marquez2016}). However, it provides higher IRX values for a given M$_{*}$ at low stellar masses (log(M$_{*}$ [M$_{\odot}$]) < 10). 

\item If our LBG sample is stacked as a function of a combination of both M$_{*}$ and $\beta_{\mathrm{UV}}$ (LBG-$M\beta$1 and LBG-$M\beta$2), the sample shows a large dispersion on the IRX-M$_{*}$ and IRX-$\beta_{\mathrm{UV}}$ planes. We find that the evolution to higher M$_{*}$ or redder $\beta_{\mathrm{UV}}$ makes them differ from the main IRX-M$_{*}$ and IRX-$\beta_{\mathrm{UV}}$ relations. New empirical IRX relation combining  $\beta_{\mathrm{UV}}$ and M$_{*}$ is suggested.
\end{itemize}

Finally, we investigate which dust attenuation curve is more likely to reproduce our LBG sample. We use \cite{Noll2009} dust attenuation law recipe and derive a variation of $\delta$ (change in the slope of the dust attenuation law with respect to Calzetti's law) from -0.3 to 0.3, which is well correlated with M$_{*}$. Steeper dust attenuation curves than Calzetti's are favored in low stellar mass LBGs ($\log(M_{*} [M_{\odot}]) < 10.25$), while grayer dust attenuation curves than Calzetti's are favored in high stellar mass LBGs ($\log(M_{*} [M_{\odot}]) > 10.25$). Combining \textbf{the simulation and stacked} results from our LBG sample, we corroborate that the slope of the dust attenuation curve is one of the main drivers to shape the IRX-$\beta_{\mathrm{UV}}$ plane.

\begin{acknowledgements}
JAM acknowledges support from the Spanish Ministry of Science, Innovation and Universities through project number ESP2017-83197. We gratefully acknowledge the contributions of the entire COSMOS collaboration consisting of more than 100 scientists. This work makes use of TOPCAT (\url{http://www.star.bristol.ac.uk/?mbt/topcat/}.).
\end{acknowledgements}


\bibliographystyle{aa}
\bibliography{Bibliography}

\begin{appendix} 

\section{Comparison of methods to derive the $\beta_{\mathrm{UV}}$ on our LBG sample}\label{beta_comparison}

\begin{figure}[h]
   \centering
   \includegraphics[width=\hsize]{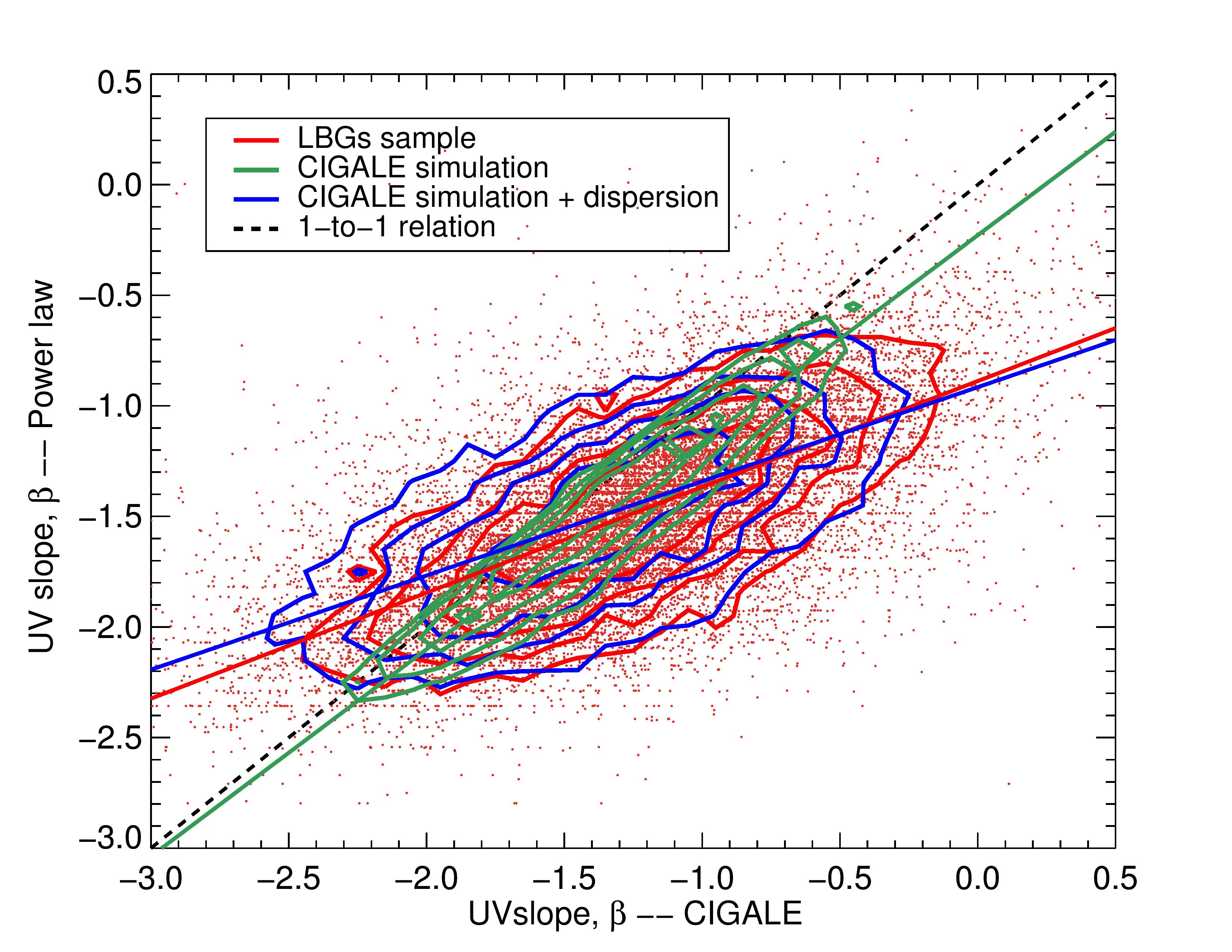}
   \caption{Comparison between the $\beta_{\mathrm{UV}}$ derived form the SED-fitting analysis and the power-law fit method. Red corresponds to the derived $\beta_{\mathrm{UV}}$ from the real data. Green represents the first simulation, where the two methods are compared using the  $\beta_{\mathrm{UV}}$ derived best-fit spectrum and the associated SED. Blue shows the second simulation,  where the two methods are compared using the  $\beta_{\mathrm{UV}}$ derived best-fit spectrum and the associated SED + scatter. Continuum color lines illustrate the linear regression for each comparative. The dashed black line is the 1-to-1 relation. } 
         \label{UVslope_simulation}
   \end{figure}

We find a large discrepancy between the $\beta_{\mathrm{UV}}$ calculations from Section \ref{slope_uvluminosity} and those presented by AM16 on the same LBG sample. Both methodologies use the same wavelength interval to perform the power-law fit. However, in Section \ref{slope_uvluminosity}, we use the best-fit synthetic spectral model obtained on the rest-frame UV-to-optical SED-fitting analysis ($\beta_{\mathrm{UV - SED}}$), and the rest-frame UV photometry  in AM16 ($\beta_{\mathrm{UV - Power}}$). Figure \ref{UVslope_simulation} shows, in red, the comparison between $\beta_{\mathrm{UV - Power}}$ (y-axis) and $\beta_{\mathrm{UV - SED}}$ (x-axis) for each LBG of the sample. This figure illustrates that the $\beta_{\mathrm{UV - Power}}$ are redder than the $\beta_{\mathrm{UV - SED}}$ in the range of $\beta_{\mathrm{UV}}$ used in that work (1.9<$\beta_{\mathrm{UV}}$<0.3) and that their differences increase at redder values. 

\cite{Finkelstein2012} used a sample of $z=4$ to 8 galaxies selected in CANDELS fields to compare the $\beta_{\mathrm{UV - SED}}$ and $\beta_{\mathrm{UV - Power}}$. They found that the results from the SED-fitting method presents essentially no bias at all redshifts, while the results from power-law method are biased toward redder values. \cite{Reddy2015} used a sample of $z\sim2$ star-forming galaxies to compare the two different $\beta_{\mathrm{UV}}$ calculations, concluding that the two measurements are highly correlated. However, Table 2 of their paper shows that the mean values of  $\beta_{\mathrm{UV - Power}}$ are 0.35 redder than the mean values of $\beta_{\mathrm{UV - SED}}$ for the same population of galaxies.

If the SED-fitting analysis provides a good model fit to the photometric data points, why would the results from the two methods be different? We perform a simulation to try to understand the origin of that differences. We perform the SED-fitting analysis over the full LBG sample to derive their best-fit spectrum and SED (photometric data points of the best-fit model). On the one hand, we use the best-fit spectrum to calculate the $\beta_{\mathrm{UV - SED}}$ associated with the SED-fitting method. On the other hand, we use the best-fit SED to obtain the $\beta_{\mathrm{UV - Power}}$ associated with the power-law method. In that case, we are working with a simulated spectrum and its associated SED, then, if both methods are consistent we should obtain similar $\beta_{\mathrm{UV}}$. Figure \ref{UVslope_simulation} shows the $\beta_{\mathrm{UV}}$ comparison, in green, when the both $\beta_{\mathrm{UV}}$ have been calculated from the same best-fit model, using the associated spectrum and SED. We can see that both methods are consistent as the differences, even if systematic, are within the uncertainty showed in Section \ref{slope_uvluminosity} ($\Delta\beta_{\mathrm{UV}}\sim0.2$).

We perform an additional test, where we include uncertainty at each photometric data point of the SED. We add a Gaussian scatter to the derived best-fit SED and recompute the $\beta_{\mathrm{UV - Power}}$. The scatter has been defined as\ equal to the observational photometric error for each band and LBG from our sample. Figure \ref{UVslope_simulation} illustrates the comparison between the $\beta_{\mathrm{UV - SED}}$ and the $\beta_{\mathrm{UV - Power}}$ derived from the best-fit model SED + scatter in blue. That shows a similar trend than that obtained when we compared the both $\beta_{\mathrm{UV}}$ in our LBG sample. This suggests that the differences between the two methods are basically due to the scatter in the photometric data points as a consequence of the photometric uncertainty.  \cite{Koprowski2018} also found similar discrepancies and suggested that these inconsistencies are driven by scatter in measured values of $\beta_{\mathrm{UV}}$ from limited photometry that serves to flatten IRX-$\beta_{\mathrm{UV}}$ relation artificially. These authors obtained that the scatter is significantly reduced by determining $\beta_{\mathrm{UV}}$ from SED-fitting analysis. 

\begin{landscape}
\section{Fluxes and uncertainties at optical and NIR wavelengths for the 30 stacked LBGs SEDs at z~3}\label{table_stak_optical}

\begin{table}[h]
\begin{adjustbox}{width=1.3\textwidth,center}
\begin{tabular}{cccccccccc}
\hline
\multicolumn{10}{c}{Stacking results at optical/NIR wavelength}\\
\hline
ID & S$_{B}$ [$\mu$Jy] & S$_{V}$ [$\mu$Jy] & S$_{r}$ [$\mu$Jy] & S$_{i}$ [$\mu$Jy] & S$_{z}$ [$\mu$Jy] & S$_{Y}$ [$\mu$Jy] & S$_{J}$ [$\mu$Jy] & S$_{H}$ [$\mu$Jy] & S$_{K_{s}}$ [$\mu$Jy] \\
\hline
LBG-$L$1 & 0.119$\pm$0.006 & 0.235$\pm$0.012 & 0.257$\pm$0.014 & 0.259$\pm$0.014 & 0.262$\pm$0.014 & 0.258$\pm$0.014 & 0.292$\pm$0.016 & 0.485$\pm$0.027 & 0.676$\pm$0.038\\
LBG-$L$2 & 0.223$\pm$0.012 & 0.416$\pm$0.022 & 0.477$\pm$0.026 & 0.520$\pm$0.029 & 0.496$\pm$0.027 & 0.536$\pm$0.030 & 0.604$\pm$0.034 & 0.943$\pm$0.053 & 1.260$\pm$0.071\\
LBG-$L$3 & 0.404$\pm$0.022 & 0.739$\pm$0.041 & 0.866$\pm$0.048 & 0.975$\pm$0.055 & 0.929$\pm$0.051 & 1.043$\pm$0.059 & 1.133$\pm$0.067 & 1.707$\pm$0.098 & 2.177$\pm$0.125\\
LBG-$L$4 & 0.732$\pm$0.050 & 1.313$\pm$0.078 & 1.554$\pm$0.094 & 1.822$\pm$0.107 & 1.690$\pm$0.099 & 1.841$\pm$0.113 & 2.027$\pm$0.129 & 2.886$\pm$0.187 & 3.603$\pm$0.230\\
\hline
LBG-$M$1 & 0.178$\pm$0.010 & 0.347$\pm$0.019 & 0.399$\pm$0.022 & 0.437$\pm$0.024 & 0.418$\pm$0.023 & 0.438$\pm$0.025 & 0.487$\pm$0.028 & 0.740$\pm$0.042 & 1.004$\pm$0.056\\
LBG-$M$2 & 0.202$\pm$0.011 & 0.398$\pm$0.022 & 0.471$\pm$0.026 & 0.510$\pm$0.030 & 0.517$\pm$0.029 & 0.579$\pm$0.034 & 0.684$\pm$0.040 & 1.126$\pm$0.065 & 1.503$\pm$0.087\\
LBG-$M$3 & 0.238$\pm$0.015 & 0.467$\pm$0.028 & 0.558$\pm$0.034 & 0.626$\pm$0.040 & 0.644$\pm$0.039 & 0.743$\pm$0.048 & 0.950$\pm$0.060 & 1.689$\pm$0.102 & 2.258$\pm$0.136\\
LBG-$M$4 & 0.247$\pm$0.020 & 0.472$\pm$0.032 & 0.573$\pm$0.039 & 0.669$\pm$0.047 & 0.650$\pm$0.044 & 0.828$\pm$0.064 & 1.073$\pm$0.075 & 2.132$\pm$0.140 & 2.970$\pm$0.187\\
LBG-$M$5 & 0.268$\pm$0.027 & 0.538$\pm$0.049 & 0.625$\pm$0.054 & 0.738$\pm$0.064 & 0.700$\pm$0.057 & 0.854$\pm$0.097 & 1.234$\pm$0.108 & 2.804$\pm$0.229 & 4.220$\pm$0.334\\
LBG-$M$6 & 0.175$\pm$0.035 & 0.399$\pm$0.059 & 0.527$\pm$0.094 & 0.650$\pm$0.109 & 0.601$\pm$0.073 & 0.670$\pm$0.200 & 1.218$\pm$0.154 & 3.066$\pm$0.393 & 5.251$\pm$0.515\\
\hline
LBG-$\beta$1 & 0.191$\pm$0.010 & 0.356$\pm$0.019 & 0.409$\pm$0.022 & 0.445$\pm$0.025 & 0.442$\pm$0.024 & 0.470$\pm$0.026 & 0.532$\pm$0.030 & 0.815$\pm$0.045 & 1.057$\pm$0.059\\
LBG-$\beta$2 & 0.149$\pm$0.008 & 0.309$\pm$0.017 & 0.366$\pm$0.020 & 0.389$\pm$0.023 & 0.454$\pm$0.026 & 0.561$\pm$0.033 & 0.680$\pm$0.042 & 1.130$\pm$0.067 & 1.476$\pm$0.087\\
LBG-$\beta$3 & 0.096$\pm$0.007 & 0.235$\pm$0.015 & 0.277$\pm$0.018 & 0.349$\pm$0.023 & 0.382$\pm$0.024 & 0.537$\pm$0.042 & 0.672$\pm$0.054 & 1.149$\pm$0.089 & 1.527$\pm$0.111\\
LBG-$\beta$4 & 0.079$\pm$0.009 & 0.214$\pm$0.020 & 0.271$\pm$0.025 & 0.391$\pm$0.042 & 0.402$\pm$0.038 & 0.538$\pm$0.075 & 0.756$\pm$0.099 & 1.513$\pm$0.184 & 2.304$\pm$0.291\\
\hline
LBG-$M\beta$1-$M_0 \beta_0$& 0.213$\pm$0.012 & 0.408$\pm$0.022 & 0.480$\pm$0.026 & 0.534$\pm$0.030 & 0.524$\pm$0.029 & 0.576$\pm$0.033 & 0.675$\pm$0.038 & 1.077$\pm$0.061 & 1.399$\pm$0.078\\
LBG-$M\beta$1-$M_0 \beta_1$& 0.140$\pm$0.008 & 0.303$\pm$0.017 & 0.366$\pm$0.021 & 0.409$\pm$0.025 & 0.465$\pm$0.027 & 0.597$\pm$0.038 & 0.754$\pm$0.047 & 1.248$\pm$0.076 & 1.609$\pm$0.097\\
LBG-$M\beta$1-$M_0 \beta_2$& 0.096$\pm$0.009 & 0.223$\pm$0.018 & 0.278$\pm$0.022 & 0.368$\pm$0.033 & 0.384$\pm$0.029 & 0.477$\pm$0.052 & 0.654$\pm$0.068 & 1.137$\pm$0.109 & 1.464$\pm$0.143\\
LBG-$M\beta$1-$M_1 \beta_0$& 0.331$\pm$0.032 & 0.598$\pm$0.050 & 0.729$\pm$0.066 & 0.860$\pm$0.077 & 0.786$\pm$0.069 & 0.812$\pm$0.092 & 1.237$\pm$0.115 & 2.654$\pm$0.244 & 4.012$\pm$0.311\\
LBG-$M\beta$1-$M_1 \beta_1$& 0.187$\pm$0.022 & 0.391$\pm$0.037 & 0.463$\pm$0.045 & 0.548$\pm$0.050 & 0.572$\pm$0.047 & 0.836$\pm$0.089 & 1.118$\pm$0.107 & 2.594$\pm$0.223 & 3.761$\pm$0.293\\
LBG-$M\beta$1-$M_1 \beta_2$& 0.098$\pm$0.014 & 0.271$\pm$0.039 & 0.364$\pm$0.047 & 0.446$\pm$0.054 & 0.512$\pm$0.057 & 0.764$\pm$0.119 & 1.188$\pm$0.154 & 2.708$\pm$0.302 & 4.560$\pm$0.495\\
\hline
LBG-$M\beta$2-$M_0 \beta_0$& 0.187$\pm$0.010 & 0.361$\pm$0.020 & 0.422$\pm$0.023 & 0.465$\pm$0.026 & 0.465$\pm$0.026 & 0.508$\pm$0.029 & 0.586$\pm$0.034 & 0.900$\pm$0.051 & 1.143$\pm$0.064\\
LBG-$M\beta$2-$M_0 \beta_1$& 0.080$\pm$0.006 & 0.205$\pm$0.014 & 0.237$\pm$0.015 & 0.317$\pm$0.020 & 0.321$\pm$0.022 & 0.450$\pm$0.044 & 0.535$\pm$0.055 & 0.753$\pm$0.069 & 0.997$\pm$0.085\\
LBG-$M\beta$2-$M_1 \beta_0$& 0.254$\pm$0.015 & 0.490$\pm$0.028 & 0.587$\pm$0.034 & 0.653$\pm$0.040 & 0.665$\pm$0.039 & 0.774$\pm$0.047 & 0.957$\pm$0.058 & 1.620$\pm$0.097 & 2.169$\pm$0.127\\
LBG-$M\beta$2-$M_1 \beta_1$& 0.126$\pm$0.010 & 0.279$\pm$0.019 & 0.344$\pm$0.024 & 0.418$\pm$0.028 & 0.472$\pm$0.031 & 0.620$\pm$0.055 & 0.782$\pm$0.060 & 1.391$\pm$0.101 & 1.743$\pm$0.127\\
LBG-$M\beta$2-$M_2 \beta_0$& 0.275$\pm$0.024 & 0.520$\pm$0.038 & 0.623$\pm$0.045 & 0.712$\pm$0.052 & 0.691$\pm$0.051 & 0.835$\pm$0.071 & 1.108$\pm$0.085 & 2.175$\pm$0.156 & 2.908$\pm$0.197\\
LBG-$M\beta$2-$M_2 \beta_1$& 0.133$\pm$0.015 & 0.299$\pm$0.030 & 0.392$\pm$0.037 & 0.504$\pm$0.057 & 0.536$\pm$0.044 & 0.780$\pm$0.090 & 1.104$\pm$0.101 & 2.227$\pm$0.172 & 3.205$\pm$0.252\\
LBG-$M\beta$2-$M_3 \beta_0$& 0.325$\pm$0.036 & 0.607$\pm$0.055 & 0.708$\pm$0.066 & 0.844$\pm$0.077 & 0.800$\pm$0.074 & 0.896$\pm$0.112 & 1.262$\pm$0.136 & 2.968$\pm$0.283 & 4.326$\pm$0.361\\
LBG-$M\beta$2-$M_3 \beta_1$& 0.129$\pm$0.022 & 0.311$\pm$0.048 & 0.385$\pm$0.055 & 0.455$\pm$0.059 & 0.464$\pm$0.051 & 0.710$\pm$0.137 & 1.141$\pm$0.150 & 2.453$\pm$0.272 & 3.594$\pm$0.364\\
LBG-$M\beta$2-$M_4 \beta_0$& 0.242$\pm$0.063 & 0.420$\pm$0.087 & 0.628$\pm$0.171 & 0.783$\pm$0.201 & 0.634$\pm$0.114 & 0.553$\pm$0.158 & 1.073$\pm$0.183 & 2.585$\pm$0.571 & 4.714$\pm$0.745\\
LBG-$M\beta$2-$M_4 \beta_1$& 0.117$\pm$0.027 & 0.398$\pm$0.079 & 0.441$\pm$0.082 & 0.530$\pm$0.084 & 0.602$\pm$0.083 & 1.013$\pm$0.150 & 1.441$\pm$0.220 & 3.511$\pm$0.450 & 5.601$\pm$0.591\\
\hline

\end{tabular}
\end{adjustbox}
\end{table}

\end{landscape}

\begin{landscape}
\section{Fluxes and uncertainties at FIR and MIR wavelength for the 30 stacked LBGs SEDs at z~3. It also includes the stacking analysis performed for AzTEC (1.1mm), and previously presented in AM16.}\label{table_stak_ir}

\begin{table}[h]

\begin{adjustbox}{width=1.3\textwidth,center}
\begin{tabular}{cccccccccccc}
\hline
\multicolumn{12}{c}{Stacking results at MIR/FIR/millimeter wavelengths}\\
\hline
ID & S$_{3.6\mu m}$ [$\mu$Jy] & S$_{4.5\mu m}$ [$\mu$Jy] & S$_{5.8\mu m}$ [$\mu$Jy]& S$_{8\mu m}$ [$\mu$Jy]& S$_{24\mu m}$ [$\mu$Jy]& S$_{100\mu m}$ [mJy] & S$_{160\mu m}$ [mJy]& S$_{250\mu m}$ [mJy]&  S$_{350\mu m}$ [mJy]& S$_{500\mu m}$ [mJy]& S$_{1.1m}$ [mJy] \\
\hline
LBG-$L$1 & 0.997$\pm$0.061 & 1.218$\pm$0.066 & 1.273$\pm$0.096 & 1.442$\pm$0.143 & 11.41$\pm$1.36 & 0.092$\pm$0.034 & 0.296$\pm$0.079 &  0.569$\pm$0.151 &  0.517$\pm$0.167 &  0.391$\pm$0.163 &       ........       \\
LBG-$L$2 & 1.670$\pm$0.102 & 1.787$\pm$0.098 & 1.880$\pm$0.134 & 1.697$\pm$0.183 & 12.05$\pm$1.53 & 0.140$\pm$0.039 & 0.289$\pm$0.097 &  0.663$\pm$0.177 &  0.786$\pm$0.203 &  0.768$\pm$0.201 &       ........       \\
LBG-$L$3 & 2.410$\pm$0.164 & 2.744$\pm$0.160 & 2.614$\pm$0.228 & 2.743$\pm$0.378 & 19.01$\pm$2.67 & 0.141$\pm$0.055 & 0.311$\pm$0.138 &  0.773$\pm$0.264 &  0.970$\pm$0.309 &  0.650$\pm$0.302 &       ........       \\
LBG-$L$4 & 4.167$\pm$0.442 & 4.649$\pm$0.408 & 5.610$\pm$0.578 & 5.878$\pm$0.837 & 26.52$\pm$6.29 & <0.432>          & 0.450$\pm$0.332 &  1.442$\pm$0.703 &  1.704$\pm$0.828 &  1.966$\pm$0.825 &       ........       \\
\hline
LBG-$M$1 & 1.217$\pm$0.078 & 1.280$\pm$0.071 & 0.934$\pm$0.107 & 0.754$\pm$0.165 & 10.12$\pm$1.46 & <0.102>          & 0.280$\pm$0.092 &  0.497$\pm$0.164 &  0.544$\pm$0.183 &  0.551$\pm$0.181 &       ........       \\
LBG-$M$2 & 1.847$\pm$0.126 & 2.224$\pm$0.126 & 2.794$\pm$0.205 & 2.241$\pm$0.314 & 21.74$\pm$2.73 & 0.198$\pm$0.058 & 0.548$\pm$0.152 &  1.355$\pm$0.268 &  1.390$\pm$0.288 &  1.001$\pm$0.268 & 0.194 $\pm$ 0.093\\
LBG-$M$3 & 3.560$\pm$0.242 & 4.096$\pm$0.234 & 4.913$\pm$0.340 & 4.192$\pm$0.529 & 40.59$\pm$5.11 & 0.269$\pm$0.072 & 0.977$\pm$0.199 &  2.157$\pm$0.402 &  2.307$\pm$0.450 &  1.902$\pm$0.415 & 0.313 $\pm$ 0.143\\
LBG-$M$4 & 6.095$\pm$0.429 & 7.268$\pm$0.443 & 8.076$\pm$0.565 & 7.698$\pm$0.807 & 60.16$\pm$7.89 & 0.417$\pm$0.130 & 1.089$\pm$0.326 &  2.944$\pm$0.652 &  3.359$\pm$0.708 &  2.446$\pm$0.637 & 0.407 $\pm$ 0.209\\
LBG-$M$5 & 8.546$\pm$0.695 & 10.49$\pm$0.66  & 12.89$\pm$0.90  & 14.86$\pm$1.36  & 98.71$\pm$13.2 & 0.533$\pm$0.168 & 2.196$\pm$0.428 &  5.527$\pm$1.075 &  6.675$\pm$1.192 &  6.267$\pm$1.113 & 1.196 $\pm$ 0.415\\
LBG-$M$6 & 15.16$\pm$1.50  & 18.47$\pm$1.64  & 23.54$\pm$2.05  & 25.46$\pm$2.76  & 185.1$\pm$29.8 & 0.992$\pm$0.258 & 3.325$\pm$0.872 &  10.91$\pm$2.39  &  12.13$\pm$2.57  &  9.732$\pm$2.216 & 2.045 $\pm$ 0.916\\
\hline
LBG-$\beta$1 & 1.282$\pm$0.075 & 1.388$\pm$0.074 & 1.160$\pm$0.090 & 1.159$\pm$0.133 & 8.511$\pm$1.13 & 0.054$\pm$0.022 & 0.092$\pm$0.047 &  0.246$\pm$0.102 &  0.400$\pm$0.123 &  0.559$\pm$0.134 &       ........       \\
LBG-$\beta$2 & 2.159$\pm$0.145 & 2.588$\pm$0.151 & 2.629$\pm$0.204 & 2.706$\pm$0.326 & 24.33$\pm$3.08 & 0.195$\pm$0.051 & 0.709$\pm$0.139 &  1.424$\pm$0.262 &  1.326$\pm$0.273 &  0.986$\pm$0.249 &       ........       \\
LBG-$\beta$3 & 3.017$\pm$0.304 & 3.869$\pm$0.320 & 4.550$\pm$0.427 & 3.759$\pm$0.758 & 43.77$\pm$6.09 & 0.252$\pm$0.111 & 1.316$\pm$0.327 &  3.527$\pm$0.628 &  3.402$\pm$0.638 &  2.724$\pm$0.576 &       ........       \\
LBG-$\beta$4 & 5.439$\pm$0.963 & 6.575$\pm$0.848 & 7.102$\pm$1.227 & 9.409$\pm$1.747 & 75.74$\pm$14.2 & 0.691$\pm$0.231 & 1.465$\pm$0.566 &  3.779$\pm$1.112 &  3.271$\pm$1.079 &  1.376$\pm$0.846 &       ........       \\
\hline
LBG-$M\beta$1-$M_0 \beta_0$& 1.820$\pm$0.108 & 2.006$\pm$0.107 & 2.063$\pm$0.137 & 1.722$\pm$0.184 & 15.36$\pm$1.86 & 0.112$\pm$0.032 & 0.380$\pm$0.087 &  0.744$\pm$0.159 &  0.733$\pm$0.178 &  0.690$\pm$0.177 &       ........       \\
LBG-$M\beta$1-$M_0 \beta_1$& 2.448$\pm$0.187 & 3.016$\pm$0.183 & 3.172$\pm$0.262 & 2.841$\pm$0.483 & 30.06$\pm$3.79 & 0.223$\pm$0.068 & 0.782$\pm$0.199 &  2.126$\pm$0.389 &  2.231$\pm$0.410 &  1.697$\pm$0.368 &       ........       \\
LBG-$M\beta$1-$M_0 \beta_2$& 2.725$\pm$0.490 & 3.312$\pm$0.413 & 4.460$\pm$0.653 & 2.845$\pm$1.278 & 46.41$\pm$8.43 & 0.237$\pm$0.188 & 1.133$\pm$0.485 &  2.987$\pm$0.866 &  2.654$\pm$0.831 &  1.618$\pm$0.736 &       ........       \\
LBG-$M\beta$1-$M_1 \beta_0$& 8.931$\pm$0.703 & 11.01$\pm$0.77  & 13.72$\pm$1.07  & 16.08$\pm$1.56  & 108.7$\pm$15.5 & 0.615$\pm$0.213 & 2.556$\pm$0.538 &  6.193$\pm$1.264 &  7.361$\pm$1.378 &  5.983$\pm$1.214 &       ........       \\
LBG-$M\beta$1-$M_1 \beta_1$& 8.272$\pm$0.686 & 9.699$\pm$0.667 & 12.73$\pm$1.01  & 10.56$\pm$1.27  & 091.5$\pm$14.0 & 0.372$\pm$0.148 & 1.644$\pm$0.527 &  4.847$\pm$1.151 &  5.538$\pm$1.309 &  4.677$\pm$1.252 &       ........       \\
LBG-$M\beta$1-$M_1 \beta_2$& 12.27$\pm$1.87  & 15.44$\pm$1.96  & 16.64$\pm$2.06  & 19.99$\pm$3.06  & 124.5$\pm$26.0 & 0.823$\pm$0.325 & 1.734$\pm$0.767 &  5.817$\pm$1.939 &  6.203$\pm$2.330 &  5.006$\pm$1.693 &       ........       \\
\hline
LBG-$M\beta$2-$M_0 \beta_0$& 1.401$\pm$0.088 & 1.499$\pm$0.081 & 1.331$\pm$0.118 & 1.046$\pm$0.168 & 10.91$\pm$1.53 & 0.076$\pm$0.032 & 0.242$\pm$0.090 &  0.559$\pm$0.157 &  0.553$\pm$0.174 &  0.460$\pm$0.173 &       ........       \\
LBG-$M\beta$2-$M_0 \beta_1$& 1.623$\pm$0.308 & 2.007$\pm$0.217 & 2.237$\pm$0.450 & 1.955$\pm$0.947 & 30.04$\pm$5.69 & <0.432>          & 0.968$\pm$0.385 &  1.851$\pm$0.647 &  1.902$\pm$0.666 &  1.533$\pm$0.595 &       ........       \\
LBG-$M\beta$2-$M_1 \beta_0$& 2.962$\pm$0.193 & 3.481$\pm$0.195 & 4.210$\pm$0.281 & 3.678$\pm$0.430 & 30.31$\pm$3.78 & 0.204$\pm$0.059 & 0.850$\pm$0.166 &  1.550$\pm$0.314 &  1.564$\pm$0.356 &  1.372$\pm$0.344 &       ........       \\
LBG-$M\beta$2-$M_1 \beta_1$& 3.344$\pm$0.368 & 3.903$\pm$0.324 & 4.454$\pm$0.522 & 3.632$\pm$0.922 & 38.24$\pm$6.38 & 0.308$\pm$0.132 & 0.941$\pm$0.372 &  2.964$\pm$0.751 &  2.847$\pm$0.751 &  2.500$\pm$0.716 &       ........       \\
LBG-$M\beta$2-$M_2 \beta_0$& 5.965$\pm$0.463 & 7.005$\pm$0.439 & 7.737$\pm$0.600 & 7.520$\pm$0.931 & 60.01$\pm$8.39 & 0.374$\pm$0.146 & 1.282$\pm$0.401 &  3.441$\pm$0.773 &  3.993$\pm$0.866 &  2.986$\pm$0.765 &       ........       \\
LBG-$M\beta$2-$M_2 \beta_1$& 6.688$\pm$0.719 & 8.004$\pm$0.681 & 9.259$\pm$0.875 & 8.350$\pm$1.167 & 66.27$\pm$12.0 & 0.500$\pm$0.211 & <1.548>         &  2.158$\pm$1.095 &  2.335$\pm$1.209 &  1.466$\pm$0.940 &       ........       \\
LBG-$M\beta$2-$M_3 \beta_0$& 8.844$\pm$0.751 & 10.42$\pm$0.69  & 12.64$\pm$1.03  & 15.80$\pm$1.52  & 102.0$\pm$14.8 & 0.666$\pm$0.211 & 2.173$\pm$0.527 &  5.536$\pm$1.408 &  6.356$\pm$1.434 &  6.292$\pm$1.327 &       ........       \\
LBG-$M\beta$2-$M_3 \beta_1$& 8.095$\pm$1.110 & 10.45$\pm$1.01  & 13.14$\pm$1.24  & 13.19$\pm$2.21  & 82.31$\pm$15.2 & <0.759>          & 1.951$\pm$0.652 &  5.790$\pm$1.412 &  7.513$\pm$1.735 &  6.588$\pm$1.607 &       ........       \\
LBG-$M\beta$2-$M_4 \beta_0$& 15.15$\pm$1.73  & 18.23$\pm$1.85  & 24.68$\pm$2.65  & 25.79$\pm$4.07  & 189.2$\pm$40.1 & 0.968$\pm$0.346 & 2.445$\pm$0.996 &  10.36$\pm$2.39  &  12.16$\pm$2.93  &  10.38$\pm$3.00  &       ........       \\
LBG-$M\beta$2-$M_4 \beta_1$& 15.23$\pm$2.29  & 19.23$\pm$2.55  & 23.59$\pm$2.77  & 27.00$\pm$3.48  & 195.9$\pm$37.4 & 0.984$\pm$0.343 & 4.434$\pm$1.276 &  11.75$\pm$3.69  &  12.48$\pm$4.06  &  8.879$\pm$3.043 &       ........       \\
\hline
\end{tabular}
\end{adjustbox}
\end{table}
\end{landscape} 

\begin{landscape}
\section{Physical parameters derived by SED-fitting analysis for the 30 stacked LBGs SEDs at z~3}\label{physical_param}

\begin{table}[h]

\begin{adjustbox}{width=1.3\textwidth,center}
\begin{tabular}{cccccccccccccc}
\hline
ID  & $\tau$ [Myr] & Age [Myr] & SFR [M$_{\odot}$yr$^{-1}$] & log(M$_{*}$ [M$_{\odot}$]) & E(B-V) & A$_{\mathrm{FUV}}$ & $\delta$ & $\beta_{\mathrm{UV}}$ & log(L$_{\mathrm{FUV}}$ [L$_{\odot}$]) & U$_{min}$ & $\gamma$ & log(L$_{\mathrm{IR}}$ [L$_{\odot}$]) & log(M$_{dust}$ [M$_{\odot}$] \\
\hline
\multicolumn{13}{c}{Stacking as a function of $L_{\mathrm{FUV}}$ (LBG-$L$)}\\
\hline
LBG-$L$1 & 147 $\pm$ 334 & 132 $\pm$ 89 & 20.6 $\pm$ 6.6 & 9.58 $\pm$ 0.11 & 0.29 $\pm$ 0.03 & 2.33 $\pm$ 0.14 & 0.18 $\pm$ 0.09 & -1.76 $\pm$ 0.19 & 10.40 $\pm$ 0.03 & 38.5 $\pm$ 10.9 & 0.032 $\pm$ 0.011 & 11.38 $\pm$ 0.05 & 7.55 $\pm$ 0.26\\
LBG-$L$2 & 271 $\pm$ 439 & 157 $\pm$ 96 & 33.6 $\pm$ 5.3 & 9.61 $\pm$ 0.10 & 0.18 $\pm$ 0.02 & 2.00 $\pm$ 0.10 & -0.08 $\pm$ 0.07 & -1.56 $\pm$ 0.19 & 10.66 $\pm$ 0.024 & 28.5 $\pm$ 13.9 & 0.028 $\pm$ 0.012 & 11.48 $\pm$ 0.04 & 7.92 $\pm$ 0.52\\
LBG-$L$3 & 369 $\pm$ 488 & 179 $\pm$ 104 & 48.6 $\pm$ 9.3 & 9.76 $\pm$ 0.10 & 0.12 $\pm$ 0.02 & 1.71 $\pm$ 0.14 & -0.23 $\pm$ 0.09 & -1.47 $\pm$ 0.19 & 10.92 $\pm$ 0.03 & 31.1 $\pm$ 13.7 & 0.033 $\pm$ 0.011 & 11.59 $\pm$ 0.06 & 7.97 $\pm$ 0.58\\
LBG-$L$4  & 180 $\pm$ 368 & 143 $\pm$ 99 & 68.5 $\pm$ 21.4 & 10.06 $\pm$ 0.11 & 0.13 $\pm$ 0.03 & 1.59 $\pm$ 0.15 & -0.15 $\pm$ 0.14 & -1.47 $\pm$ 0.19 & 11.17 $\pm$ 0.03 & 19.2 $\pm$ 15.6 & 0.024 $\pm$ 0.013 & 11.80 $\pm$ 0.08 & 8.79 $\pm$ 0.72 \\
\hline
\multicolumn{13}{c}{Stacking as a function of $\beta_{\mathrm{UV}}$ (LBG-$\beta$)}\\
\hline
LBG-$\beta$1 & 443 $\pm$ 513 & 229 $\pm$ 118 & 21.0 $\pm$ 4.3 & 9.49 $\pm$ 0.09 & 0.12 $\pm$ 0.02 & 1.64 $\pm$ 0.13 & -0.22 $\pm$ 0.10 & -1.52 $\pm$ 0.19 & 10.60 $\pm$ 0.03 & 5.7 $\pm$ 7.2 & 0.034 $\pm$ 0.010 & 11.23 $\pm$ 0.06 & 8.62 $\pm$ 0.43\\
LBG-$\beta$2 & 405 $\pm$ 495 & 159 $\pm$ 94 & 63.6 $\pm$ 9.5 & 9.75 $\pm$ 0.13 & 0.24 $\pm$ 0.03 & 2.87 $\pm$ 0.14 & -0.10 $\pm$ 0.10 & -1.01 $\pm$ 0.20 & 10.54 $\pm$ 0.03 & 41.1 $\pm$ 9.3 & 0.029 $\pm$ 0.012 & 11.76 $\pm$ 0.04 & 7.89 $\pm$ 0.13\\
LBG-$\beta$3 & 250 $\pm$ 420 & 110 $\pm$ 53 & 115 $\pm$ 12 & 9.97 $\pm$ 0.08 & 0.37 $\pm$ 0.02 & 3.83 $\pm$ 0.19 & -0.02 $\pm$ 0.04 & -0.67 $\pm$ 0.22 & 10.43 $\pm$ 0.03 & 35.2 $\pm$ 10.9 & 0.014 $\pm$ 0.007 & 12.06 $\pm$ 0.03 & 8.34 $\pm$ 0.19\\
LBG-$\beta$4 & 279 $\pm$ 450 & 216 $\pm$ 180 & 118 $\pm$ 48 & 10.42 $\pm$ 0.18 & 0.40Â¡ $\pm$ 0.05 & 4.11 $\pm$ 0.28 & -0.001 $\pm$ 0.095 & -0.46 $\pm$ 0.26 & 10.42 $\pm$  0.03 & 41.6 $\pm$ 9.5 & 0.036 $\pm$ 0.009 & 12.18 $\pm$ 0.08 & 8.28 $\pm$ 0.17\\
\hline
\multicolumn{13}{c}{Stacking as a function of stellar mass (LBG-$M$)}\\
\hline
LBG-$M$1 & 465 $\pm$ 514 & 112 $\pm$ 48 & 29.6 $\pm$ 3.4 & 9.23 $\pm$ 0.10 & 0.13 $\pm$ 0.02 & 1.89 $\pm$ 0.10 & -0.28 $\pm$ 0.07 & -1.60 $\pm$ 0.19 & 10.59 $\pm$ 0.03 & 30.9 $\pm$ 14.2 & 0.026 $\pm$ 0.013 & 11.33 $\pm$ 0.04 & 7.77 $\pm$ 0.68 \\
LBG-$M$2 & 404 $\pm$ 494 & 122 $\pm$ 54 & 64.2 $\pm$ 7.4 & 9.64 $\pm$ 0.10 & 0.22 $\pm$ 0.02 & 2.61 $\pm$ 0.13 & -0.09 $\pm$ 0.04 & -1.36 $\pm$ 0.19 & 10.65 $\pm$ 0.03 & 36.9 $\pm$ 10.6 & 0.026 $\pm$ 0.012 & 11.74 $\pm$ 0.03 & 7.94 $\pm$ 0.16\\
LBG-$M$3 & 484 $\pm$ 520 & 249 $\pm$ 112 & 88.1 $\pm$ 10.3 & 10.12 $\pm$ 0.10 & 0.28 $\pm$ 0.02 & 2.84 $\pm$ 0.14 & 0.01 $\pm$ 0.05 & -1.17$\pm$ 0.20 & 10.72 $\pm$ 0.03 & 35.2 $\pm$ 10.7 & 0.025 $\pm$ 0.012 & 11.95 $\pm$ 0.03 & 8.18 $\pm$ 0.16\\
LBG-$M$4 & 344 $\pm$ 461 & 382 $\pm$ 251 & 99.8 $\pm$ 27.1 & 10.54 $\pm$ 0.11 & 0.35 $\pm$ 0.04 & 3.13 $\pm$ 0.21 & 0.10 $\pm$ 0.09 & -1.04 $\pm$ 0.21 & 10.73 $\pm$ 0.03 & 33.0 $\pm$ 11.1 & 0.029 $\pm$ 0.012 & 12.10 $\pm$ 0.06 & 8.35 $\pm$ 0.17\\
LBG-$M$5 & 443 $\pm$ 525 & 414 $\pm$ 270 & 182 $\pm$ 58 & 10.81 $\pm$ 0.11 & 0.44 $\pm$ 0.04 & 3.67 $\pm$ 0.21 & 0.16 $\pm$ 0.08 & -1.08 $\pm$ 0.24 & 10.78 $\pm$ 0.03 & 24.8 $\pm$  9.6 & 0.017 $\pm$ 0.009 & 12.39 $\pm$ 0.05 & 8.82 $\pm$ 0.18\\
LBG-$M$6 & 444 $\pm$ 523 & 612 $\pm$ 451 & 257 $\pm$ 95 & 11.22 $\pm$ 0.10 & 0.55 $\pm$ 0.05 & 4.44 $\pm$ 0.25 & 0.18 $\pm$ 0.07 & -0.75 $\pm$ 0.33& 10.66 $\pm$ 0.04 & 25.7 $\pm$ 10.5 & 0.018 $\pm$ 0.009 & 12.62 $\pm$ 0.05 & 9.04 $\pm$ 0.20\\
\hline
\multicolumn{13}{c}{Stacking as a function of ($\beta_{\mathrm{UV}}$, M$_{*}$) - LBG-$M\beta$1}\\
\hline
LBG-$M\beta$1-$M_0 \beta_0$ & 422 $\pm$ 502 & 200 $\pm$ 106 & 37.9 $\pm$ 5.9 & 9.66 $\pm$ 0.10 & 0.17 $\pm$ 0.02 & 2.08 $\pm$ 0.12 & -0.12 $\pm$ 0.08 & -1.40 $\pm$ 0.19 & 10.66 $\pm$ 0.03 & 36.2 $\pm$ 11.6 & 0.031 $\pm$ 0.011 & 11.52 $\pm$ 0.05 & 7.73 $\pm$ 0.24\\
LBG-$M\beta$1-$M_0 \beta_1$ & 413 $\pm$ 497 & 117 $\pm$ 45 & 89.4 $\pm$ 8.9 & 9.77 $\pm$ 0.09 & 0.27 $\pm$ 0.02 & 3.19 $\pm$ 0.16 & -0.10 $\pm$ 0.03 & -0.87 $\pm$ 0.20 & 10.54 $\pm$ 0.03 & 35.3 $\pm$ 10.9 & 0.018 $\pm$ 0.009 & 11.90 $\pm$ 0.03 & 8.16 $\pm$ 0.19\\
LBG-$M\beta$1-$M_0 \beta_2$ & 346 $\pm$ 475 & 104 $\pm$ 57 & 114 $\pm$ 20 & 9.87 $\pm$ 0.15 & 0.34 $\pm$ 0.03 & 3.77 $\pm$ 0.19 & -0.05 $\pm$ 0.06 & -0.70 $\pm$ 0.23 & 10.43 $\pm$ 0.03 & 37.3 $\pm$ 11.59 & 0.024 $\pm$ 0.012 & 12.02 $\pm$ 0.05 & 8.25 $\pm$ 0.31\\
LBG-$M\beta$1-$M_1 \beta_0$ & 368 $\pm$ 487 & 428 $\pm$ 308 & 171 $\pm$ 66 & 10.91 $\pm$ 0.11 & 0.48 $\pm$ 0.04 & 3.53 $\pm$ 0.20 & 0.25 $\pm$ 0.08 & -1.26 $\pm$ 0.23 & 10.83 $\pm$ 0.03 & 31.8 $\pm$ 11.4 & 0.019 $\pm$ 0.009 & 12.42 $\pm$ 0.06 & 8.73 $\pm$ 0.23\\
LBG-$M\beta$1-$M_1 \beta_1$ & 580 $\pm$ 558 & 473 $\pm$ 273 & 159 $\pm$ 36 & 10.71 $\pm$ 0.11 & 0.39 $\pm$ 0.03 & 3.78 $\pm$ 0.19 & 0.05 $\pm$ 0.07 & -0.73 $\pm$ 0.25 & 10.64 $\pm$ 0.03 & 27.1 $\pm$ 12.3 & 0.017 $\pm$ 0.009 & 12.29 $\pm$ 0.05 & 8.75 $\pm$ 0.35\\
LBG-$M\beta$1-$M_1 \beta_2$ & 308 $\pm$ 472 & 457 $\pm$ 390 & 132 $\pm$ 65 & 11.00 $\pm$ 0.11 & 0.45 $\pm$ 0.04 & 4.23 $\pm$ 0.21 & 0.07 $\pm$ 0.10 & -0.23 $\pm$ 0.31 & 10.51 $\pm$ 0.04 & 27.7 $\pm$ 13.7 & 0.029 $\pm$ 0.012 & 12.39 $\pm$ 0.04 & 8.85 $\pm$ 0.56\\
\hline
\multicolumn{13}{c}{Stacking as a function of ($\beta_{\mathrm{UV}}$, M$_{*}$) - LBG-$M\beta$2}\\
\hline
LBG-$M\beta$2-$M_0 \beta_0$ & 377 $\pm$ 490 & 148 $\pm$ 69 & 28.8 $\pm$ 3.3 & 9.42 $\pm$ 0.07 & 0.13 $\pm$ 0.01 & 1.89 $\pm$ 0.09 & -0.29 $\pm$ 0.04 & -1.42 $\pm$ 0.19 & 10.60 $\pm$ 0.02 & 33.7 $\pm$ 12.9 & 0.028 $\pm$ 0.011 & 11.35 $\pm$ 0.03 & 7.67 $\pm$ 0.48\\
LBG-$M\beta$2-$M_0 \beta_1$ & 237 $\pm$ 410 & 92 $\pm$ 47 & 61.4 $\pm$ 17.5 & 9.61 $\pm$ 0.11 & 0.28 $\pm$ 0.04 & 3.26 $\pm$ 0.20 & -0.09 $\pm$ 0.07 & -0.78 $\pm$ 0.22 & 10.37 $\pm$ 0.03 & 29.3 $\pm$ 14.9 & 0.024 $\pm$ 0.012 & 11.75 $\pm$ 0.08 & 8.28 $\pm$ 0.75\\
LBG-$M\beta$2-$M_1 \beta_0$ & 270 $\pm$ 440 & 155 $\pm$ 94 & 69.6 $\pm$ 13.5 & 9.95 $\pm$ 0.11 & 0.23 $\pm$ 0.02 & 2.58 $\pm$ 0.12 & -0.06 $\pm$ 0.08 & -1.21 $\pm$ 0.19 & 10.74 $\pm$ 0.03 & 38.5 $\pm$ 10.5 & 0.028 $\pm$ 0.012 & 11.84 $\pm$ 0.04 & 8.02 $\pm$ 0.18\\
LBG-$M\beta$2-$M_1 \beta_1$ & 329 $\pm$ 468 & 166 $\pm$ 107 & 93.6 $\pm$ 24.6 & 10.01 $\pm$ 0.15 & 0.31 $\pm$ 0.03 & 3.37 $\pm$ 0.20 & -0.05 $\pm$ 0.07 & -0.71 $\pm$ 0.21 & 10.52 $\pm$ 0.03 & 29.4 $\pm$ 13.7 & 0.021 $\pm$ 0.012 & 11.97 $\pm$ 0.08 & 8.41 $\pm$ 0.50\\ 
LBG-$M\beta$2-$M_2 \beta_0$ & 449 $\pm$ 521 & 351 $\pm$ 230 & 122 $\pm$ 31 & 10.49 $\pm$ 0.13 & 0.36 $\pm$ 0.04 & 3.18 $\pm$ 0.21 & 0.10 $\pm$ 0.08 & -1.14 $\pm$ 0.22 & 10.77 $\pm$ 0.03 & 32.3 $\pm$ 12.2 & 0.022 $\pm$ 0.011 & 12.16 $\pm$ 0.06 & 8.47 $\pm$ 0.27\\
LBG-$M\beta$2-$M_2 \beta_1$ & 343 $\pm$ 481 & 401 $\pm$ 320 & 91.9 $\pm$ 33.4 & 10.56 $\pm$ 0.10 & 0.33 $\pm$ 0.04 & 3.48 $\pm$ 0.26 & -0.04 $\pm$ 0.09 & -0.42 $\pm$ 0.25 & 10.56 $\pm$ 0.03 & 37.1 $\pm$ 12.6 & 0.034 $\pm$ 0.001 & 12.09 $\pm$ 0.08 & 8.34 $\pm$ 0.66\\
LBG-$M\beta$2-$M_3 \beta_0$ & 378 $\pm$ 473 & 483 $\pm$ 304 & 155 $\pm$ 46 & 10.88 $\pm$ 0.07 & 0.44 $\pm$ 0.04 & 3.38 $\pm$ 0.17 & 0.22 $\pm$ 0.08 & -1.19 $\pm$ 0.24 & 10.83 $\pm$ 0.03 & 25.7 $\pm$ 12.9 & 0.024 $\pm$ 0.012 & 12.35 $\pm$ 0.04 & 8.84 $\pm$ 0.41\\
LBG-$M\beta$2-$M_3 \beta_1$ & 463 $\pm$ 533 & 418 $\pm$ 286 & 175 $\pm$ 54 & 10.78 $\pm$ 0.13 & 0.45 $\pm$ 0.04 & 4.20 $\pm$ 0.21 & 0.07 $\pm$ 0.08 & -0.51 $\pm$ 0.35 & 10.53 $\pm$ 0.04 & 23.3 $\pm$ 12.5 & 0.017 $\pm$ 0.009 & 12.37 $\pm$ 0.05 & 8.93 $\pm$ 0.39\\
LBG-$M\beta$2-$M_4 \beta_0$ & 555 $\pm$ 542 & 932 $\pm$ 571 & 235 $\pm$ 83 & 11.36 $\pm$ 0.11 & 0.58 $\pm$ 0.05 & 4.18 $\pm$ 0.26 & 0.27 $\pm$ 0.09 & -1.01 $\pm$ 0.40 & 10.69 $\pm$ 0.06 & 18.9 $\pm$ 11.7 & 0.021 $\pm$ 0.011 & 12.60 $\pm$ 0.05 & 9.27 $\pm$ 0.43\\
LBG-$M\beta$2-$M_4 \beta_1$ & 407 $\pm$ 511 & 416 $\pm$ 296 & 303 $\pm$ 111 & 11.10 $\pm$ 0.13 & 0.50 $\pm$ 0.05 & 4.72 $\pm$ 0.23 & 0.06 $\pm$ 0.09 & -0.28 $\pm$ 0.39 & 10.61 $\pm$ 0.05 & 33.5 $\pm$ 12.1 & 0.019 $\pm$ 0.010 & 12.65 $\pm$ 0.05 & 8.97 $\pm$ 0.35\\
\hline
\end{tabular}
\end{adjustbox}
\end{table}

\end{landscape} 

\end{appendix}

\end{document}